\documentclass[prd,preprint,nofootinbib,12pt]{revtex4}
\usepackage{amsfonts}
\usepackage{amsmath}
\usepackage{amssymb}
\usepackage{graphicx}
\usepackage{hyperref}
\usepackage{xcolor}

\newcommand{\dd}{{\rm{d}}} 

\newcommand{\im}{\mathrm{i}}

\begin{document}

\title{New class of rotating charged black holes \\
with nonaligned electromagnetic field}

\author{Hryhorii Ovcharenko}
\email{hryhorii.ovcharenko@matfyz.cuni.cz}
\affiliation{Charles University, Faculty of Mathematics and Physics,
Institute of Theoretical Physics,
V~Hole\v{s}ovi\v{c}k\'ach 2, 18000 Prague 8, Czechia}

\author{Ji\v{r}\'i Podolsk\'{y}}
\email{jiri.podolsky@matfyz.cuni.cz}
\affiliation{Charles University, Faculty of Mathematics and Physics,
Institute of Theoretical Physics,
V~Hole\v{s}ovi\v{c}k\'ach 2, 18000 Prague 8, Czechia}

\begin{abstract}
    We present a large family of twisting and expanding solutions to the Einstein-Maxwell equations of algebraic type D, for which the two double principal null directions (PNDs) of the Weyl tensor are not aligned with the null eigendirections of the Faraday tensor. In addition to systematically deriving this new class, we present its various metric forms and convenient parameterizations. We show that in Boyer-Lindquist-type coordinates these solutions depend on 7 parameters, namely the Kerr and NUT (Newman-Unti-Tamburino) twist parameters $a$ and $l$, mass parameter $m$, acceleration $\alpha$, strength of the Maxwell field  $|c|$, and angular parameters $\beta, \gamma$ that represent two duality rotations of the Faraday tensor, which include the rotation between the electric and magnetic charges generating the aligned part of the Maxwell field. This coordinate parameterization, analogous to the Griffiths-Podolsk\'y form of the Pleba\'nski-Demia\'nski solutions, allows us to perform various limits, explicitly identify the subcases, and determine the physical interpretation of the new class. Interestingly, by considering the limit with no acceleration ($\alpha\to 0$), one obtains either the famous Kerr-Newman-NUT black holes (if the parameter $|c|$ remains constant) or the novel Kerr-Bertotti-Robinson black holes, announced recently in our work [Kerr Black Hole in a Uniform Bertotti-Robinson Magnetic Field: An Exact Solution, Phys. Rev. Lett. {\bf 135} (2025) 18, 181401] (if $|c|\rightarrow \infty$ while $\alpha |c|=\mathrm{const.}$). We may thus conclude that this new class of spacetimes represents twisting charged accelerating black holes, immersed in an external magnetic (or electric) field. In the non-twisting subcase, we obtain the previously known solution of Van den Bergh-Carminati..
\end{abstract}

\date{\today}

\maketitle

\tableofcontents

\section{Introduction}

Since the discovery and understanding of the Schwarzschild solution, there has been a great interest in exact solutions describing black holes. Different possible generalizations of the Schwarzschild spacetime, describing charged (Reissner-Nordstr\"om) and rotating (Kerr and Kerr-Newman) black holes were found (for their reviews see, e.g., \cite{Stephani:2003tm, GriffithsPodolsky:2009}). All these solutions are of special physical interest because they describe unique asymptotically flat spacetimes. Indeed, these are \textit{the only asymptotically flat} black hole solutions to the Einstein-Maxwell equations with specific symmetries (according to the uniqueness theorems). In addition, these solutions are of algebraic type~D, and the two null eigendirections of the electromagnetic field are aligned with both the (double-degenerate) PNDs of the Weyl tensor. Up to the moment when the Kerr and the Kerr-Newman solutions appeared in the 1960s, some other solutions were also known that satisfied all the properties mentioned above, except the global asymptotical flatness (namely the C-metric \cite{levi1918ds2} and the Taub-NUT solution \cite{Newman1963}). This initiated (along with purely mathematical interest) attempts to find \textit{all} solutions satisfying the aforementioned properties except of the asymptotical flatness. An extensive work on this topic was done to fully integrate the Einstein-Maxwell equations for the case of type~D spacetimes with a completely aligned electromagnetic field \cite{Carter1968,Kinnersley1969,Debever1969,Debever1971} (in \cite{Debever1979,Debever1981} this class was denoted as the $\mathcal{D}$~class). The most general such solution was found in \cite{Debever1983,Debever1984}. However, due to its great complexity, it was not physically fully interpreted. Nevertheless, it was shown that the only expanding spacetimes within this class are contained in the Pleba\'nski-Demia\'nski solution \cite{Plebanski1976} (see also its revised versions \cite{GRIFFITHS2006,Griffiths2005,PodolskyGriffiths:2006,Podolsk2021,Podolsk2023} and very recent analysis \cite{Ovcharenko2024, Ovcharenko2025}).

Along with the idea of violating the asymptotical flatness condition, a question of what happens if, in addition, one \textit{does not require the alignment condition} of all the PNDs of the Faraday and Weyl tensors. The search was primarily focused on the case such that only one of the eigendirections of the Faraday tensor $F_{\alpha\beta}$ is not aligned with the PNDs of the Weyl tensor $C_{\alpha\beta\gamma\delta}$. A strong interest in such a setup was initiated by the Kundt-Tr\"{u}mper theorem \cite{kundt1963beitrage}, which states that if the special condition ${3\Psi_2=\pm 2|\Phi_1|^2}$ for the Newman-Penrose scalars is not satisfied, then (at least one of) the optical scalars $\kappa$ and $\sigma$ associated with the aligned PND congruence must be zero \cite{Stephani:2003tm}. This condition implies a great simplification of the field equations, and it allowed to extend the already known solutions of vacuum field equations with geodesic and shear-free PND \cite{Robinson1969} to the case where only one eigendirection of the Faraday tensor is aligned \cite{Robinson1969_2} (note that the original solutions are more general and do not require a spacetime to be of algebraic type~D). Later, it was shown \cite{Leroy1978, Leroy1979} that \textit{if in addition one assumes the solution to be of algebraic type D, then the corresponding spacetime is flat in the vacuum case}. This means that this spacetime does not contain a Kerr black hole as a subcase, and thus cannot be considered as a proper non-aligned analog of the Kerr-Newman spacetime.

One may attempt to take one step further and consider \textit{fully non-aligned spacetimes} (such that \emph{none} of the eigendirections of the Faraday tensor is aligned with \textit{any} of the two double-degenerate PNDs of the Weyl tensor). Unfortunately, for a long time such a general setup was not of interest because of the great complexity of the corresponding field equations. Relatively recently, Van den Bergh \cite{VandenBergh2016} in his work proved that \textit{non-aligned algebraic type~D solutions of the Einstein-Maxwell equations are not compatible with non-zero cosmological constant~$\Lambda$}. This important observation simplified the search for corresponding solutions. Subsequently, a non-aligned solution was explicitly found by Van den Bergh and Carminati \cite{VandenBergh2020}. It was obtained under the assumption of \emph{zero twist}, and in its presented form it is complicated. Moreover, it does not directly allow to obtain various limiting subcases.

Our current work aims to overcome these issues, and to generalize the solution found in \cite{VandenBergh2020} to the twisting situation. Instead of presenting the solution in just one mathematical form, our study also elaborates on presenting various metric forms and the corresponding parameterizations. This is useful because it allows us to give a basic physical interpretation of the new class of metrics, and to take various limits. Our final form of the solution, see Eq.~(\ref{GP_metr}) below, depends on 7 different parameters: mass $m$, Kerr-like rotation $a$, NUT-like twist $l$, acceleration $\alpha$, magnitude of the electromagnetic field $|c|$, and two duality rotation parameters $\gamma$ and $\beta$. By considering special cases we show that the electromagnetic field can be split into two parts: the field generated by charges of a black hole itself, and the external electromagnetic field (that becomes uniform for $m=0$). Thus the whole solution has an interpretation as a \textit{charged twisting black hole in an external magnetic (or electric) field.} Moreover, this parametrization allows us to show that in the static limit, we recover the non-twisting solution previously found in \cite{VandenBergh2020}.

Despite the fact that the solutions we are considering here are not asymptotically flat \emph{globally} (because of the presence of an external electromagnetic field, which backreacts on the geometry), models describing \emph{black holes in magnetic fields} were always of great interest. There exists a wide class of Melvin-type spacetimes representing black holes immersed into a Bonnor-Melvin background \cite{Bonnor54,Melvin64}. These solutions were found by employing the Harrison transformation to the Schwarzschild, Kerr, and Kerr-Newman black holes \cite{Ernst1976, Ernst1976_2} (see also recent developments \cite{Klemm25}). They were used to investigate the influence of the magnetic fields on the motion of uncharged \cite{Galtsov78, Esteban84, Bizyaev24} and charged \cite{Dadhich79} particles, the black hole shadows \cite{Wang21}, the image of a black hole created by accretion discs \cite{Hou22}, the thermodynamical properties \cite{Gibbons2014}, and other effcts. However, there are several objections against the applicability of these spacetimes as realistic models of magnetized black holes. Both charged and uncharged particles cannot escape to infinity in the equatorial plane \cite{Galtsov78}. Even more, geodesics may become chaotic in this spacetime \cite{Karas92}, ergoregions may extend to infinity \cite{Gibbons13}, and these spacetimes are of algebraic type I \cite{Pravda05}, which means that they generally radiate. Our solution does not belong to the Melvin-type spacetimes. It may thus overcome these issues, and become a more useful playground for astrophysical investigations and various studies of mathematical relativity.

Our new solution may also become interesting in the context of \emph{supergravity}. It is known that the Einstein-Maxwell theory in 4 spacetime dimensions corresponds to the bosonic part of the ${D=4}$, ${\mathcal{N}=2}$ (ungauged) minimal supergravity \cite{Freedman1977}. The solution to the whole supergravity (with a fermionic sector) may be obtained from any spacetime of the Einstein-Maxwell theory if in such a spacetime there exists \textit{a Killing spinor} (for an explanation see \cite{Gibbons1982}). Moreover, such a solution preserves at least a quarter of the supersymmetries (this represents the so-called BPS states). Imposing the condition of the existence of the Killing spinor within the Pleba\'nski-Demia\'nski class of solutions gave rise to supersymmetric type D spacetimes \cite{AlonsoAlberca2000,Maeda2011,Klemm2013,Crisafio:2024fyc}. The case of non-aligned electromagnetic field presented here was not considered in this context yet.

\newpage
The paper is organized as follows. In Sec.~II we establish an ansatz for the metric and electromagnetic field, along with writing the field equations in the Newman-Penrose formalism. In Sec.~III we solve them, and present the most general solution. In Secs.~IV-V we elaborate on finding the Pleba\'{n}ski-Demia\'{n}ski-type form, adding the acceleration and twist parameters to it. In Sec. VI we investigate the structure of the non-aligned electromagnetic field. In Sec.~VII we present the Griffiths-Podolsk\'{y}-type form resembling the Boyer-Lindquist coordinates. Sec.~VIII investigates several important subcases that allow us to interpret the new class. In Sec.~IX we give a summary of various particular cases, and present the physical interpretation for each parameter, appearing in the new class. In Sec.~X we analyze positions of horizons, cosmic strings, and the electromagnetic field for the most general case. Concluding remarks are given in Sec.~XI.

\newpage

\section{Motivation and general setup}

As was outlined in the Introduction, our work aims to find a class of twisting solutions of type D with a fully non-aligned electromagnetic field $F_{\alpha\beta}$. For the integration, it is natural to choose a null tetrad aligned with the the PNDs of the Weyl tensor $C_{\alpha\beta\gamma\delta}$. This means that the only non-zero Weyl scalar is $\Psi_2$, whereas all scalars $\Phi_0, \Phi_1,\Phi_2$ of the electromagnetic field are nonzero in general.

Let us formulate an ansatz for deriving our new class. We expect that it will reduce to the non-twisting metric obtained by Van den Bergh and Carminati in \cite{VandenBergh2020}, namely Eq.~(85) therein. With a trivial renaming ${t \mapsto \eta}$, ${x \mapsto q}$, ${s \mapsto p}$, ${z \mapsto \sigma}$, ${k \mapsto K}$ it has the form
\begin{align} \label{VandenBergh-metric}
    \dd s^2=\dfrac{K^2}{2N}\,\big(\!-e\,\xi^2 \dd \eta^2+e\,\xi^{-2} \dd q^2
           +\varsigma^{-2}\dd p^2+\varsigma^2 \dd \sigma^2\big)\,,
\end{align}
where ${e=\pm 1}$, $\xi^2$ is the quartic function of $q$ (Eqs.~(83) and (101) in \cite{VandenBergh2020}), $\varsigma^2$ is the quartic function of $p$ (Eqs. (84) and (102) in \cite{VandenBergh2020}), and the function $N$ is quadratic in both  $p$ and $q$ (Eqs.~(82) and (103) in \cite{VandenBergh2020}). By simple redefinitions $Q=e\, \xi^2\,,~ P=\varsigma^2\,,~ \Omega^2=2N/K^2\,,$ this metric becomes
\begin{align}
    \dd s^2=\dfrac{1}{\Omega^2}\Big(-Q \,\dd \eta^2+\dfrac{\dd q^2}{Q}+\dfrac{\dd p^2}{P}+P \,\dd \sigma^2\Big),
    \label{orig_metr}
\end{align}
where $P(p)$ and $Q(q)$ are quartic functions of $p$ and $q$, respectively, while the conformal factor $\Omega^2(p,q)$ is quadratic in both $p$ and $q$. The metric (\ref{orig_metr}) resembles the C-metric that represents accelerating black holes \cite{Griffiths2006_2} (see also Eq.~(14.6) in \cite{GriffithsPodolsky:2009}). The basic difference between the solution of \cite{VandenBergh2020} and the C-metric  (along with the fact that the electromagnetic field is no longer aligned) is that for the C-metric the function $\Omega^2$ is given by a simple expression ${\Omega=1-pq}$, while for the Van den Bergh-Carminati solution it is much more complicated (given by Eq.~(82) and (103) in \cite{VandenBergh2020}).

It is well-known that twisting generalization of the C-metric is the Pleba\'nski-Demia\'nski solution,
which in the original form \cite{Plebanski1976} reads
\begin{align}
    \dd s^2=\frac{1}{\Omega^2}\Big[-\dfrac{Q}{\rho^2}(\dd \eta-p^2\dd\sigma)^2+\dfrac{\rho^2}{Q}\dd q^2
    +\dfrac{\rho^2}{P}\dd p^2+\dfrac{P}{\rho^2}(\dd \eta+q^2\dd \sigma)^2\Big],
    \label{metr}
\end{align}
where
\begin{align}\label{def-rho-OmegaPD}
    \rho^2=p^2+q^2\,, \quad\hbox{and}\quad \Omega=1-pq\,.
\end{align}

The solution (\ref{orig_metr}) basically differs by a more complicated conformal factor~$\Omega$, and by the absence of $\rho^2$, so the basic idea is that a \emph{twisting generalization} of the Van den Bergh-Carminati solution could be given by the metric (\ref{metr}), but with more general metric functions  $P(p)$, $Q(q)$, and $\Omega(p,q)$. This is indeed so: Explicit expressions for these metric functions, namely \eqref{P-PD}, \eqref{Q-PD}, \eqref{Omega-C'} with \eqref{metr_shift}, \eqref{rho-shift}, will be found below by applying the field equations.

For the integration of the field equations, we will employ the Newman-Penrose formalism. The natural orthonormal tetrad for the metric (\ref{metr}) is
\begin{align}\label{orthogonal-tetrad}
    \mathbf{e}_0&=\sqrt{\dfrac{\Omega^2}{Q \rho^2}}\,(q^2\partial_{\eta}-\partial_{\sigma})\,,\qquad
    \mathbf{e}_1 =\sqrt{\dfrac{\Omega^2 Q}{\rho^2}}\,\partial_q\,,\nonumber\\
    \mathbf{e}_2&=\sqrt{\dfrac{\Omega^2P}{\rho^2}}\,\partial_p\,,\qquad
    \mathbf{e}_3 =\sqrt{\dfrac{\Omega^2}{P \rho^2}}\,(p^2\partial_{\eta}+\partial_{\sigma})\,.
\end{align}
The corresponding null tetrad ${\mathbf{k}=\tfrac{1}{\sqrt{2}}\,(\mathbf{e}_0+\mathbf{e}_1)}$, ${\mathbf{l}=\tfrac{1}{\sqrt{2}}\,(\mathbf{e}_0-\mathbf{e}_1)}$, ${\mathbf{m}=\tfrac{1}{\sqrt{2}}\,(\mathbf{e}_3+\im\,\mathbf{e}_2)}$ reads
\begin{align}
    \mathbf{k}&=\sqrt{\dfrac{\Omega^2}{2Q\rho^2}}\,(q^2\partial_{\eta}-\partial_{\sigma}+Q\partial_q)\,,\nonumber\\
    \mathbf{l}&=\sqrt{\dfrac{\Omega^2}{2Q\rho^2}}\,(q^2\partial_{\eta}-\partial_{\sigma}-Q\partial_q)\,,\nonumber\\
    \mathbf{m}&=\sqrt{\dfrac{\Omega^2}{2P\rho^2}}\,(p^2\partial_{\eta}+\partial_{\sigma}+\im\, P\partial_p)\,,\nonumber\\
    \bar{\mathbf{m}}&=\sqrt{\dfrac{\Omega^2}{2P\rho^2}}\,(p^2\partial_{\eta}+\partial_{\sigma}-\im\,P\partial_p)\,.\label{null_tetr}
\end{align}

One can now check that the metric ansatz (\ref{metr}) is of algebraic type D. We start by calculating the spin coefficients associated with the null vector fields $\mathbf{k}$ and $\mathbf{l}$, namely
\begin{align}
    \rho_{\rm sc}&=\mu=\dfrac{1}{2}\sqrt{\dfrac{Q \Omega^2}{2\rho^2}}\Bigg[\Big(\ln \dfrac{\Omega^2}{\rho^2}\Big)_{,q}+\im\, (\ln\rho^2)_{,p}\Bigg],\nonumber\\
    \tau&=\pi=\dfrac{1}{2}\sqrt{\dfrac{P\Omega^2}{2\rho^2}}\Bigg[(\ln \rho^2)_{,q}+\im\, \Big(\ln \dfrac{\Omega^2}{\rho^2}\Big)_{,p}\Bigg],\nonumber\\
    \alpha&=\beta=\dfrac{1}{4}\sqrt{\dfrac{P\Omega^2}{2\rho^2}}\Bigg[(\ln \rho^2)_{,q}+\im\, \Big(\ln \dfrac{P}{\rho^2\Omega^2}\Big)_{,p}\Bigg],\nonumber\\
    \epsilon&=\gamma=\dfrac{1}{4}\sqrt{\dfrac{Q\Omega^2}{2\rho^2}}\Bigg[ \Big(\ln \dfrac{Q}{\rho^2\Omega^2}\Big)_{,q}+\im\,(\ln \rho^2)_{,p}\Bigg],\nonumber\\
    \kappa&=\nu=0,\qquad\sigma=\lambda=0\,.\label{opt_sc_5}
\end{align}

The Ricci tensor components are given by expressions
\begin{align}
        \Phi_{00}&=\dfrac{Q\Omega}{2\rho^2}\,\Omega_{,qq}=\Phi_{22},\label{Phi_00_02}\\
        \Phi_{02}&=-\dfrac{P\Omega}{2\rho^2}\,\Omega_{,pp}=\bar{\Phi}_{20},\label{Phi_02_20}\\
        \Phi_{01}&=\dfrac{\sqrt{P Q}}{2\rho^4}\Omega\Big(\im\,\rho^2 \,\Omega_{,qp}-(q+\im\,p)(\Omega_{,q}+\im\,\Omega_{,p})\Big)  =\Phi_{21} \label{Phi_01}\\
        &=\bar{\Phi}_{10}=\bar{\Phi}_{12}\,,\label{Phi_12}\\
        \Phi_{11}&=\dfrac{1}{8\rho^4}\Bigg[\rho^6\Omega^4\Big(\dfrac{Q_{,q}}{\Omega^2\rho^4}\Big)_{,q}
             +2Q\Big[\big(\Omega^2(\rho^2)_{,q}\big)_{,q}-\rho^2\Omega\Omega_{,qq}\Big]\nonumber\\
    &\hspace{10mm} -\rho^6\Omega^4\Big(\dfrac{P_{,p}}{\Omega^2\rho^4}\Big)_{,p}
             -2P\Big[\big(\Omega^2(\rho^2)_{,p}\big)_{,p}-\rho^2\Omega\Omega_{,pp}\Big]\Bigg],\label{phi_11}
\end{align}
and the Ricci scalar reads
\begin{align}
    R&=-\dfrac{\Omega^2}{\rho^2}\Bigg[\Big[P\Big(\ln\dfrac{P}{\Omega^3}\Big)_{,p}\Big]_{,p}+\Big[Q\Big(\ln\dfrac{Q}{\Omega^3}\Big)_{,q}\Big]_{,q} \nonumber\\
    &\hspace{12mm} +\dfrac{3}{2}\Big[P\big((\ln\Omega^2)_{,p}\big)^2+Q\big((\ln\Omega^2)_{,q}\big)^2\Big]\Bigg]. \label{R_expr}
\end{align}
The only non-zero Weyl scalar is
\begin{align}
    \Psi_2=\dfrac{\Omega^2}{12}\dfrac{(q+\im\,p)^2}{q-\im\,p}\Bigg[\Big(\dfrac{Q}{(q+\im\,p)^3}\Big)_{,qq}
    +\Big(\dfrac{P}{(q+\im\,p)^3}\Big)_{,pp}\Bigg],
\end{align}
confirming that the metric (\ref{metr}) is indeed of algebraic type~D.

Now let us discuss the corresponding electromagnetic field. As the spacetime (\ref{metr}) admits two Killing vectors, namely $\partial_{\eta}$ and $\partial_\sigma$, the electromagnetic field also has to be invariant with respect to translations along them. This means that the vector potential $A_{\mu}$ has to be independent of the coordinates $\eta$ and $\sigma$. Then, by performing a convenient gauge fixing, it is possible to write the electromagnetic field potential in the form.
\begin{equation}
    \mathbf{A}=A_{\eta}(q,p)\,\dd\eta+A_{\sigma}(q,p)\,\dd\sigma.\label{A_mu_expr}
\end{equation}
Calculating the Faraday tensor components from (\ref{A_mu_expr}), it turns out that the relation
\begin{equation}
    \Phi_2=\Phi_0  \label{phi2=phi0}
\end{equation}
holds.

This ``symmetry condition'' will be crucial for solving the Einstein-Maxwell field equations
\begin{align}
    \Phi_{ij}&=2\Phi_i\bar{\Phi}_j\,, \label{EMeqs}\\
    R&=0\,, \label{R=0}
\end{align}
along with the Maxwell equations
\begin{align}
    D\Phi_1-\bar{\delta}\Phi_0 &=(\pi-2\alpha)\Phi_0+2\rho\Phi_1-\kappa \Phi_2\,,\nonumber\\
    D\Phi_2-\bar{\delta}\Phi_1 &=-\lambda \Phi_0+2\pi \Phi_1+(\rho-2\varepsilon)\Phi_2\,, \nonumber\\
    \Delta\Phi_0-\delta\Phi_1  &=(2\gamma-\mu)\Phi_0-2\tau \Phi_1+\sigma \Phi_2\,,\nonumber\\
    \Delta\Phi_1-\delta\Phi_2  &=\nu \Phi_0-2\mu\Phi_1+(2\beta-\tau)\Phi_2\,.\label{Max_eqs}
\end{align}
By solving them we obtain the real metric functions ${P, Q, \Omega^2}$ and the electromagnetic field represented by the complex NP scalars ${\Phi_0=\Phi_2}$ and ${\Phi_1}$.

\section{Solving the field equations}

Now we are ready to solve the Einstein and Maxwell equations. We start with the equations (\ref{EMeqs}) for $\Phi_{00}$ and $\Phi_{02}$, given by \eqref{Phi_00_02} and \eqref{Phi_02_20}, namely
\begin{align}
    \Phi_{00}&= \dfrac{Q\Omega}{2\rho^2}\,\Omega_{,qq}=2\Phi_0\bar{\Phi}_0\,,\label{phi_00}\\
    \Phi_{02}&=-\dfrac{P\Omega}{2\rho^2}\,\Omega_{,pp}=2\Phi_0\bar{\Phi}_2\,.\label{phi_02}
\end{align}
A fully general solution is difficult to obtain, so we will make a further assumption. In analogy to the Van den Bergh-Carminati work \cite{VandenBergh2020} (in the spirit of the Newman-Janish algorithm) we will \emph{assume} that $\Phi_0$ (equal to $\Phi_2$ due to (\ref{phi2=phi0})) has the form
\begin{align}
    \Phi_0&=\dfrac{c'}{\Omega}\,\dfrac{\sqrt{PQ}}{q+\im\,p}=\Phi_2\,,\label{phi_0}
\end{align}
where $c'$ is a complex constant (here we use a primed symbol to avoid confusion with the rescaled constant $c$ introduced later in (\ref{rescl_params})).

This form of ${\Phi_0=\Phi_2}$ allows us to obtain an analytical solution for $\Omega$. Indeed, rewriting Eqs. (\ref{phi_00}) and (\ref{phi_02}) as
\begin{align}
    Q\,\Omega_{,qq}
     = 4\dfrac{c'\bar{c}'}{\Omega^3}\,PQ
     = -P\,\Omega_{,pp} \,,\label{om_eq_2}
\end{align}
one can solve these equations to obtain two alternative forms of~$\Omega$, namely\footnote{Notice that ${Q\,\Omega_{,qq} + P\,\Omega_{,pp} =0}$ admits a trivial solution ${\Omega_{,qq} = 0 = \Omega_{,pp}}$ for which $\Omega$ is  \emph{linear both} in $p$ and $q$. This leads to the Pleba\'nski-Demia\'nski class of type D solutions with aligned electromagnetic field such that ${\Omega=1-pq}$, see \eqref{def-rho-OmegaPD}.}
\begin{align}
     \dfrac{4 c'\bar{c}'P}{f_1(p)}+f_1(p)\big[q+f_2(p)\big]^2=\Omega^2=-\dfrac{4 c'\bar{c}'Q}{f_3(q)}+f_3(q)\big[p+f_4(q)\big]^2\,.\label{Omega2_PQ}
\end{align}

The functions $f_1, f_2, f_3, f_4$ can be found explicitly. First of all, we note that the left-hand-side of \eqref{Omega2_PQ} is quadratic in~$q$, whereas the right-hand-side is quadratic in~$p$. Therefore, taking $\dfrac{\partial^5\Omega^2}{\partial p^3\partial q^2}$ of (\ref{Omega2_PQ}) it follows that
\begin{align}
    (f_1)_{,ppp}=0 \qquad \Rightarrow\qquad  f_1=f_{10}+f_{11}\,p+f_{12}\,p^2,
\end{align}
where $f_{10},~f_{11},~f_{12}$ are just constants. If one takes the partial derivatives in the reversed order, $\dfrac{\partial^5\Omega^2}{\partial q^3\partial p^2}$, one analogously obtains
\begin{align}
    (f_3)_{,qqq}=0 \qquad \Rightarrow\qquad  f_3=f_{30}+f_{31}\,q+f_{32}\,q^2.
\end{align}

Similarly, by comparing the left-hand-side and the right-hand-side of $\dfrac{\partial^4\Omega^2}{\partial p^3\partial q}$ of \eqref{Omega2_PQ}, one obtains
\begin{align}
    (f_1 f_2)_{,ppp}=0 \qquad \Rightarrow\qquad     f_2=\dfrac{f_{20}+f_{21}\,p+f_{22}\,p^2}{f_1(p)},
    \label{f1f2}
\end{align}
and by considering $\dfrac{\partial^4\Omega^2}{\partial q^3\partial p}$ one gets
\begin{align}
    (f_3 f_4)_{,qqq}=0 \qquad \Rightarrow\qquad     f_4=\dfrac{f_{40}+f_{41}\,q+f_{42}\,q^2}{f_3(q)},
    \label{f3f4}
\end{align}

Moreover, from $(\Omega^2)_{,ppp}$ it follows that
\begin{align}
    \Big(\dfrac{4c'\bar{c}'P}{f_1}+f_1 f_2^2\Big)_{,ppp} = 0\,.
\end{align}
Integrating this equation and using the expression \eqref{f1f2}, one obtains
\begin{align}
    f_1^{-1}\big[4c'\bar{c}'P+(f_{20}+f_{21}\,p+f_{22}\,p^2)^2\big] = P_0+P_1\, p+P_2 \,p^2,
    \label{auxiliar}
\end{align}
where $P_0,~P_1,~P_2$ are also constants. So that
\begin{align}
    P=\dfrac{1}{4c'\bar{c}'}\Big[(f_{10}+f_{11}\,p+f_{12}\,p^2)(P_0+P_1\, p+P_2 \,p^2)-(f_{20}+f_{21}\,p+f_{22}\,p^2)^2\Big].
\end{align}
From this expression it follows that \emph{the function $P$ is at most quartic in $p$}. Analogously, from $(\Omega^2)_{,qqq}$  one finds that \emph{the function $Q$ is at most quartic in $q$}.

Finally, the function $\Omega^2$ can be written using the left-hand-side of \eqref{Omega2_PQ}, \eqref{f1f2} and \eqref{auxiliar}
 in the form
\begin{align}
    \Omega^2&= f_1^{-1}\big(4c'\bar{c}'P+f_1^2 f_2^2\big)+2f_1f_2 \,q+f_1 \,q^2\nonumber\\
    & =(P_0+P_1\, p+P_2 \,p^2) + 2(f_{20}+f_{21}\,p+f_{22}\,p^2)\,q + (f_{10}+f_{11}\,p+f_{12}\,p^2)\,q^2,
\end{align}
so that the function $\Omega^2$ is a \textit{quadratic function in both coordinates $p$ and $q$}. The same conclusion follows, of course, from the right-hand-side of \eqref{Omega2_PQ}.

To summarize, by solving the Einstein-Maxwell field equations for $\Phi_{00}$ and $\Phi_{02}$ we obtained general forms of the metric functions
\begin{align}
    P&=\hat{a}_0+\hat{a}_1\,p+\hat{a}_2\,p^2+\hat{a}_3\,p^3+\hat{a}_4\,p^4, \label{p_exp_1}\\ Q&=\hat{b}_0+\hat{b}_1\,q+\hat{b}_2\,q^2+\hat{b}_3\,q^3+\hat{b}_4\,q^4, \label{q_exp_1}\\
    \Omega^2&=\big(c_{00}+c_{01}q+c_{02}q^2\big)+\big(c_{10}+c_{11}q+c_{12}q^2\big)p+\big(c_{20}+c_{21}q+c_{22}q^2\big)p^2 ,\label{omega_exp_1}
\end{align}
where $\hat{a}_i, \hat{b}_i, c_{ij}$ are constants.

Due to \eqref{Phi_00_02}  and \eqref{phi2=phi0}, i.e. ${\Phi_2=\Phi_0}$, also the field equation for $\Phi_{22}$ is now satisfied. In order to solve the remaining equations \eqref{EMeqs} for  $\Phi_{01}, \Phi_{11}, \Phi_{12}$ (${=\Phi_{10}}$) and ${R=0}$, we have to determine the electromagnetic field component $\Phi_1$. It is obtained from the Maxwell equations (\ref{Max_eqs}). In view of \eqref{opt_sc_5} for the spin coefficients, they simplify to
\begin{align}
    D\Phi_1-\bar{\delta}\Phi_0 &= (\pi-2\alpha)\Phi_0+2\rho\Phi_1 \,,\label{Max_eqs-simplified1}\\
    D\Phi_0-\bar{\delta}\Phi_1 &= 2\pi \Phi_1+(\rho-2\varepsilon)\Phi_0\,, \label{Max_eqs-simplified2}\\
    -\Delta\Phi_0+\delta\Phi_1 &= 2\pi \Phi_1+(\rho-2\varepsilon)\Phi_0 \,,\label{Max_eqs-simplified3}\\
    -\Delta\Phi_1+\delta\Phi_0 &= (\pi-2\alpha)\Phi_0+2\rho\Phi_1\,,\label{Max_eqs-simplified4}
\end{align}
where ${D=k^{\mu}\,\nabla_{\mu}}$, ${\delta=m^{\mu}\,\nabla_{\mu}}$, ${\Delta=l^{\mu}\,\nabla_{\mu}}$. For the null tetrad \eqref{null_tetr} we get the relations
\begin{align}
    -\Delta \Phi_{i}=D \Phi_{i}\,,\qquad \delta\Phi_{i}=-\bar{\delta}\Phi_{i}\,.
\end{align}
Indeed, using the fact that $\Phi_i$ depend only on the coordinates $q$ and $p$ we have
\begin{align}
    -\Delta \Phi_i = \sqrt{\dfrac{Q\Omega^2}{2\rho^2}}(\Phi_i)_{,q} = D \Phi_{i}\,,\qquad
    \delta \Phi_i = \im\,\sqrt{\dfrac{P\Omega^2}{2\rho^2}}(\Phi_i)_{,p} = -\bar{\delta} \Phi_{i}\,.
\end{align}
It is thus obvious that the Maxwell equation \eqref{Max_eqs-simplified3} is identical to \eqref{Max_eqs-simplified2}, and the Maxwell equation \eqref{Max_eqs-simplified4} is identical to \eqref{Max_eqs-simplified1}. These two Maxwell equations have the explicit form
\begin{align}
    \sqrt{\dfrac{Q\Omega^2}{2\rho^2}}(\Phi_1)_{,q}+\im\,\sqrt{\dfrac{P\Omega^2}{2\rho^2}}(\Phi_0)_{,p}&=\dfrac{\im}{2}\sqrt{\dfrac{P\Omega^2}{2\rho^2}}\Big( \ln\dfrac{\Omega^4}{P}\Big)_{,p}\Phi_0
    +\sqrt{\dfrac{Q\Omega^2}{2\rho^2}}\Big[\Big(\ln \dfrac{\Omega^2}{\rho^2}\Big)_{,q}+\im\,( \ln \rho^2)_{,p}\Big]\Phi_1,\\
    \sqrt{\dfrac{Q\Omega^2}{2\rho^2}}(\Phi_0)_{,q}+\im\,\sqrt{\dfrac{P\Omega^2}{2\rho^2}}( \Phi_1)_{,p}&=\sqrt{\dfrac{P\Omega^2}{2\rho^2}}\Big[(\ln \rho^2)_{,q}+\im\,\Big(\ln \dfrac{\Omega^2}{\rho^2}\Big)_{,p}\Big]\Phi_1
    +\dfrac{1}{2}\sqrt{\dfrac{Q \Omega^2}{2\rho^2}}\Big( \ln\dfrac{\Omega^4}{Q}\Big)_{,q}\Phi_0,
\end{align}
which can be rewritten compactly as
\begin{align}
    &\sqrt{Q}\Bigg[\dfrac{\Omega^2}{\rho^2}\Big(\dfrac{\rho^2}{\Omega^2}\Phi_1\Big)_{,q}-\im\,(\ln \rho^2)_{,p}\, \Phi_1\Bigg]
    +\im\,\Omega^2\Big(\dfrac{\sqrt{P}}{\Omega^2}\Phi_0\Big)_{,p}=0\,,\\
    &\sqrt{P}\Bigg[\dfrac{\Omega^2}{\rho^2}\Big(\dfrac{\rho^2}{\Omega^2}\Phi_1\Big)_{,p}+\im\,(\ln \rho^2)_{,q}\, \Phi_1\Bigg]
    -\im\,\Omega^2\Big(\dfrac{\sqrt{Q}}{\Omega^2}\Phi_0\Big)_{,q}=0\,.
\end{align}
For the assumed form (\ref{phi_0}) of $\Phi_0$ and applying the field equations (\ref{om_eq_2}), we obtain the set of equations
\begin{align}
    &\dfrac{\Omega^2}{\rho^2}\Big(\dfrac{\rho^2}{\Omega^2}\Phi_1\Big)_{,q}-\im\,(\ln \rho^2)_{,p}\, \Phi_1+\im\,\Omega^2\Big(\dfrac{1}{4\bar{c}'}\dfrac{\Omega_{,qq}}{q+\im\,p}\Big)_{,p}=0\,,\\
    &\dfrac{\Omega^2}{\rho^2}\Big(\dfrac{\rho^2}{\Omega^2}\Phi_1\Big)_{,p}+\im\,(\ln \rho^2)_{,q}\, \Phi_1+\im\,\Omega^2\Big(\dfrac{1}{4\bar{c}'}\dfrac{\Omega_{,pp}}{q+ip}\Big)_{,q}=0\,.
\end{align}
This can be further simplified by making a substitution
\begin{equation}
    \Phi_1=\dfrac{1}{4\bar{c}'}\dfrac{\Omega^2}{(q+\im\,p)^2}\,g(p,q)\,,
\end{equation}
where $g$ is a general function of both $p$ and $q$. This brings the Maxwell equations to
\begin{align}
    &\Big[\,g-     p^2\Big(\dfrac{\Omega_{,q}}{p}\Big)_{,p}\,\Big]_{,q}
    +\im\,q\,\Omega_{,qqp}=0\,, \label{max_eq_sim_1}\\
    &\Big[\,g+\im\,q^2\Big(\dfrac{\Omega_{,p}}{q}\Big)_{,q}\,\Big]_{,p}
    -     p\,\Omega_{,ppq}=0\,,\label{max_eq_sim_2}
\end{align}
which allow us to easily find the function $g$ as
\begin{equation}
    g=g_0+p^2\Big(\dfrac{\Omega_{,q}}{p}\Big)_{,p}-\im\,q^2\Big(\dfrac{\Omega_{,p}}{q}\Big)_{,q}
    \,,
\end{equation}
where $g_0$ is a complex constant (this can be directly checked by inserting  this $g$ into the equations (\ref{max_eq_sim_1}), (\ref{max_eq_sim_2})). The electromagnetic field is thus given by the components
\begin{align}
    \Phi_0&=\Phi_2=\dfrac{c'}{\Omega}\,\dfrac{\sqrt{PQ}}{q+\im\,p}\,, \label{phi0z}\\
    \Phi_1&=\dfrac{1}{4\bar{c}'}\,\dfrac{\Omega^2}{(q+\im\,p)^2}\Big[\,g_0+p^2\Big(\dfrac{\Omega_{,q}}{p}\Big)_{,p}
    -\im\,q^2\Big(\dfrac{\Omega_{,p}}{q}\Big)_{,q}\Big], \label{phi1z}
\end{align}
where $c'$ and $g_0$ are any complex constants.

Next we solve the Einstein-Maxwell equation ${{\Phi_{01}=2\Phi_0\bar{\Phi}_1}}$. The Ricci tensor component $\Phi_{01}$ is given by (\ref{Phi_01}), while the right-hand side reads
\begin{align}
    2\Phi_0\bar{\Phi}_1=\dfrac{\sqrt{PQ}}{2\rho^4}\Omega\Big[\bar{g}_0(q+\im\,p)+\im\,\rho^2\,\Omega_{,qp}-(q+\im\, p)(\Omega_{,q}+\im\,\Omega_{,p})\Big].
\end{align}
It follows that in \eqref{phi1z} we have to set
\begin{align}
    g_0=0 \,.
\end{align}

It can now be proven that the Maxwell field is generated by the specific potential \eqref{A_mu_expr}, written as the 1-form ${\mathbf{A}=A_{\mu}\,\dd x^\mu}$, as
\begin{align}
    \mathbf{A} = \dfrac{1}{4\bar{c}'}\Big[\,
    \dfrac{\Omega_{,q}-\im\,\Omega_{,p}}{q+\im\,p}\,\dd\eta
    +\Big(\dfrac{q^2\,\Omega_{,q}+\im\,p^2\,\Omega_{,p}}{q+\im\,p}-\Omega\Big)\,\dd\sigma
    \Big].   \label{A-1}
\end{align}
This complex expression gives the Faraday 2-form ${\mathbf{F} :=\dd\mathbf{A}}$. Defining
${\mathbf{F}^* := \mathbf{F}+\im\,\tilde{\mathbf{F}}}$, where  ${\tilde{\mathbf{F}}}$ is the Hodge dual, the \emph{energy-momentum tensor} of the field is ${T_{\alpha\beta}=\tfrac{1}{2}\,F_{\,\alpha}^{*\,\gamma}\bar{F}^*_{\,\beta \gamma}}$ (see Sec.~5.2 in \cite{Stephani:2003tm}). Also the corresponding real counterpart can be obtained from the relation
\begin{align}
    \mathbf{A}^{\rm real}=2\,\mathrm{Re} \mathbf{A}.
\end{align}
Moreover, there is an \emph{invariant} ${F^*_{\,\mu\nu}\,F^{*\,\mu\nu} = 16 (\Phi_0 \Phi_2-\Phi_1^2)}$, distinguishing \emph{null} and \emph{non-null} fields. Since ${\Phi_0=\Phi_2\ne\Phi_1}$ in our case, we conclude that the electromagnetic field is \emph{generally non-null}.

Finally, we explicitly find the metric functions (\ref{p_exp_1})--(\ref{omega_exp_1}) by applying the last two field equations, namely ${R=0}$, where $R$ is a Ricci scalar given by \eqref{R_expr}, and ${\Phi_{11}=2\Phi_1\bar{\Phi}_{1}}$, where $\Phi_{11}$ is given by (\ref{phi_11}) and
\begin{equation}
    2\Phi_1\bar{\Phi}_{1}=\dfrac{1}{c'\bar{c}'}\dfrac{\Omega^4}{8\rho^4}
    \Bigg[\Big[p^2\Big(\dfrac{\Omega_{,q}}{p}\Big)_{,p}\,\Big]^2
         +\Big[q^2\Big(\dfrac{\Omega_{,p}}{q}\Big)_{,q}\,\Big]^2\Bigg].
    \label{phi_11_em}
\end{equation}
A direct calculation shows that the explicit solution to all the Einstein-Maxwell equations has the form
\begin{align}
    \hat{a}_0&= \dfrac{4c_{02}-c_{01}^2}{16c'\bar{c}'}\,, \nonumber\\
    \hat{a}_1&= \dfrac{c_{01}+c_{02}c_{10}+c_{12}}{4c'\bar{c}'}\,, \nonumber\\
    \hat{a}_2&=-\dfrac{2+2c_{02}^2-2c_{10}c_{12}+c_{01}c_{21}-2c_{22}}{8c'\bar{c}'}\,,\label{a-hat-i}\\
    \hat{a}_3&= \dfrac{c_{21}+c_{22}c_{10}-c_{02}c_{12}}{4c'\bar{c}'}\,, \nonumber\\
    \hat{a}_4&=-\dfrac{4c_{02}c_{22}+c_{21}^2}{16c'\bar{c}'}\,, \nonumber
\end{align}
and
\begin{align}
    \hat{b}_0&= \dfrac{4c_{02}+c_{10}^2}{16c'\bar{c}'}\,, \nonumber\\
    \hat{b}_1&=-\dfrac{c_{10}-c_{02}c_{01}+c_{21}}{4c'\bar{c}'}\,, \nonumber\\
    \hat{b}_2&= \dfrac{2+2c_{02}^2+c_{10}c_{12}-2c_{01}c_{21}-2c_{22}}{8c'\bar{c}'}\,,\label{b-hat-i}\\
    \hat{b}_3&=-\dfrac{c_{12}+c_{22}c_{01}+c_{02}c_{21}}{4c'\bar{c}'}\,, \nonumber\\
    \hat{b}_4&= -\dfrac{4c_{02}c_{22}-c_{12}^2}{16c'\bar{c}'}\,, \nonumber
\end{align}
where
\begin{equation}\label{c-ij}
    c_{00}=1\,,\qquad
    c_{11}=-2\,,\qquad
    c_{20}=-c_{02}\,.
\end{equation}
There are thus 6 free real parameters $c_{01}, c_{10}, c_{02}, c_{12}, c_{21}, c_{22}$, in addition to the complex charge parameter $c'$.

Let us remark that ${c_{00}=1}$ can \emph{always} be obtained (when ${c_{00}\ne0}$) by performing a constant rescaling of the conformal factor ${\Omega\mapsto S\,\Omega}$, accompanied by the rescailing of the charge parameter ${c'\mapsto S^2 c'}$ which keeps the Einstein-Maxwell field equations unchanged. Moreover, the metric \eqref{metr} and the electromagnetic field (\ref{phi0z}), (\ref{phi1z}) are unchanged under a constant rescaling of the coordinates ${p\mapsto cp}$, ${q\mapsto cq}$, ${\eta\mapsto\eta/c}$, ${\sigma\mapsto  \sigma/c^3}$, ${P\mapsto c^4 P}$, ${Q\mapsto c^4 Q}$, ${c'\mapsto c'/c^3}$. This allows us to scale one of the coefficients $c_{ij}$ to an arbitrary value, for example ${c_{11}=-2}$.

Our next task is to get a clear physical interpretation of this new large class of exact solutions given by \eqref{metr}, \eqref{def-rho-OmegaPD} and (\ref{phi0z}), (\ref{phi1z}) with  \eqref{a-hat-i}--\eqref{c-ij}. In particular, first we have to identify the subclass of well-known spacetimes of algebraic type~D, namely the Kerr-Newman-NUT black holes and their accelerating (C-metric) generalizations. To this end we have to find an explicit relation to their Pleba\'nski-Demia\'nski (PD) form \cite{Plebanski1976} and the Griffiths-Podolsk\'y (GP) form \cite{GRIFFITHS2006,Podolsk2021,Podolsk2023,Ovcharenko2024,Ovcharenko2025}.

\newpage

\section{The Pleba\'nski-Demia\'nski-type form}\label{sec_PD}

The relations \eqref{a-hat-i}--\eqref{c-ij} are complicated, and physical meaning of the coefficients $c_{ij}$ is not obvious at all. It turns out that a considerable simplification, and subsequent understanding, is achieved if we perform a shift of the coordinates
\begin{equation}\label{pq-shift}
    p\mapsto p+p_0\,,\qquad  q\mapsto q+q_0 \,,
\end{equation}
where $p_0, q_0$ are some constants. This allows us to impose an additional (gauge) condition
\begin{equation}
    c_{01}=0=c_{10}\,.
\end{equation}
Basically, by (\ref{pq-shift}) we are replacing the two real parameters $c_{01}, c_{10}$ by the two equivalent real parameters $p_0, q_0$.

After the shift \eqref{pq-shift} the metric is not given by (\ref{metr}), but has a slightly more general form
\begin{align}
    \dd s^2=\frac{1}{\Omega^2}\Big[-\dfrac{Q}{\rho^2}\big[\dd\eta-(p+p_0)^2\dd \sigma\big]^2
    +\dfrac{\rho^2}{Q}\dd q^2+\dfrac{\rho^2}{P}\dd p^2+\dfrac{P}{\rho^2}\big[\dd\eta+(q+q_0)^2\dd\sigma\big]^2\Big],
    \label{metr_shift}
\end{align}
where
\begin{equation}\label{rho-shift}
\rho^2=(p+p_0)^2+(q+q_0)^2\,,
\end{equation}
and the conformal factor is
\begin{equation}\label{Omega_c_ij}
    \Omega^2 = 1 - 2pq + c_{02}(q^2-p^2) + (c_{21}\,p + c_{12}\,q)\,pq + c_{22}\,p^2q^2\,.
\end{equation}
The electromagnetic field now reads
\begin{align}
 \Phi_0&=\Phi_2=\dfrac{c'}{\Omega}\,\dfrac{\sqrt{PQ}}{(q+q_0)+\im\, (p+p_0)}\,,\label{phi_0_shift}\\
 \Phi_1&=\dfrac{1}{4\bar{c}'}\,\dfrac{\Omega^2}{[(q+q_0)+\im\, (p+p_0)]^2}
    \Big[(p+p_0)^2\Big(\dfrac{\Omega_{,q}}{p+p_0}\Big)_{,p}-\im\,(q+q_0)^2\Big(\dfrac{\Omega_{,p}}{q+q_0}\Big)_{,q}\,\Big],\label{phi_1_shift}
\end{align}
arising from the shifted 1-form potential \eqref{A-1},
\begin{align}
    \mathbf{A} = \dfrac{1}{4\bar{c}'}\Big[\,
    \dfrac{\Omega_{,q}-\im\,\Omega_{,p}}{(q+q_0)+\im\,(p+p_0)}\,\dd\eta
    +\Big(\dfrac{(q+q_0)^2\,\Omega_{,q}+\im\,(p+p_0)^2\,\Omega_{,p}}{(q+q_0)+\im\,(p+p_0)}-\Omega\Big)\,\dd\sigma
    \Big].   \label{A-2}
\end{align}

The functions $P$ and $Q$ \emph{remain quartic} in $p$ and $q$, respectively, with the coefficients $a'_i, b'_i$,
\begin{align}
    P&=a'_0+a'_1\,p+a'_2\,p^2+a'_3\,p^3+a'_4\,p^4, \label{P-PD}\\
    Q&=b'_0+b'_1\,q+b'_2\,q^2+b'_3\,q^3+b'_4\,q^4. \label{Q-PD}
\end{align}
Solving the field equations ${R=0}$ and ${\Phi_{11}=2\Phi_1\bar{\Phi}_{1}}$, one obtains that the coefficients $a'_i, b'_i$  in \eqref{P-PD}, \eqref{Q-PD} are related to the coefficients $c_{ij}$ in \eqref{Omega_c_ij} by expressions analogous to \eqref{a-hat-i}, \eqref{b-hat-i}, but now with ${c_{01}=0=c_{10}}$ (and $\hat{a}_i, \hat{b}_i$ renamed to $a'_i, b'_i$).

All the constants $a'_i, b'_i$ are thus determined by 4~real constants $c_{02}, c_{12}, c_{21}, c_{22}$, and by the complex charge $c'$. However, instead of $c_{02}, c_{12}, c_{21}, c_{22}$, it is now convenient to introduce 4~new auxiliary parameters defined by the expressions
\begin{align}
    k'&:= \dfrac{c_{02}}{4c'\bar{c}'}\,, \nonumber\\
    \epsilon'&:=\dfrac{1+c_{02}^2-c_{22}}{4c'\bar{c}'}\,, \nonumber\\
    m'&:= \dfrac{2c_{21}-c_{02}c_{12}}{16c'\bar{c}'}\,, \nonumber\\
    n'&:= \dfrac{2c_{12}+c_{02}c_{21}}{16c'\bar{c}'}\,. \label{def-PD-coeff}
\end{align}
Then the coefficients in \eqref{P-PD} and \eqref{Q-PD} take compact forms
\begin{align}
    a'_0&= k'\,, \nonumber\\
    a'_1&= 2\,\dfrac{n'-C'm'}{1+C'^2}\,, \nonumber\\
    a'_2&=-\epsilon' \,,\label{a'-i}\\
    a'_3&= 2\,\dfrac{(1+2C'^2)m'-C'n'}{1+C'^2}\,, \nonumber\\
    a'_4&=-(1+4C'^2)k' + 2C' \Big(\epsilon'-\dfrac{(m'+C'n')^2}{k'(1+C'^2)^2}\Big)\,, \nonumber
\end{align}
and
\begin{align}
    b'_0&= k'\,, \nonumber\\
    b'_1&=-2\,\dfrac{m'+C'n'}{1+C'^2}\,, \nonumber\\
    b'_2&= \epsilon' \,,\label{b'-i}\\
    b'_3&=-2\,\dfrac{(1+2C'^2)n'+C'm'}{1+C'^2}\,, \nonumber\\
    b'_4&=-(1+4C'^2)k' + 2C' \Big(\epsilon'+\dfrac{(n'-C'm')^2}{k'(1+C'^2)^2}\Big)\,, \nonumber
\end{align}
where
\begin{equation}
  C' := \dfrac{c_{02}}{2} = 2c'\bar{c}'k'\,.
    \label{C' definition}
\end{equation}
Moreover, the conformal factor $\Omega^2$ becomes
\begin{align}
    \Omega^2 = (1-pq)^2& +2C'(q^2-p^2)+4(C'^2-c'\bar{c}'\epsilon')\, p^2q^2\nonumber\\
         & +\dfrac{8c'\bar{c}'}{1+C'^2}\big[(m'+C'n')p+(n'-C'm')q\big]\,pq\,. \label{Omega-C'}
\end{align}

The Einstein-Maxwell field equations also impose an \emph{additional constraint} on the shift coefficients $p_0, q_0$, namely
\begin{align}
  (\,p_0-A')^2  + (q_0-B')^2 = R'^2\,,\label{p0q0_cond}
\end{align}
where the constants $A', B', R'$ are
\begin{align}
 A'=&\dfrac{2C'(n'-C'm')}{(1+C'^2)[(1+4C'^2)k'-2C'\epsilon']}\,,\nonumber\\
 B'=&\dfrac{2C'(m'+C'n')}{(1+C'^2)[(1+4C'^2)k'-2C'\epsilon']}\,,\label{defAB}\\  R'=&\dfrac{2C'\sqrt{m'^2+n'^2}}{\sqrt{1+C'^2}\,[(1+4C'^2)k'-2C'\epsilon']}\,.\nonumber
\end{align}
This represents a \emph{circle} of radius $R'$ in the space of the coefficients ($p_0,q_0$). The constraint \eqref{p0q0_cond} can thus be naturally \textit{parametrized by a single angle $\beta$} as
\begin{align}
 p_0&= A' + R' \sin\beta\,,\nonumber\\
 q_0&= B' + R' \cos\beta\,.\label{p0q0-in-ABR}
\end{align}
Instead of the two coefficients $p_0, q_0$ there is, in fact, \textit{only one new independent parameter}~$\beta$.
Interestingly, ${p_0=0=q_0}$ can \emph{always} be achieved for the \emph{particular choice} of $\beta$ such that
\begin{align}
    \sin \beta_0=-\dfrac{A'}{R'}=\dfrac{-(n'-C'\,m')}{\sqrt{1+C'^2}\,\sqrt{m'^2+n'^2}}
    \quad \Rightarrow \quad
    \cos \beta_0=-\dfrac{B'}{R'}=\dfrac{-(m'+C'n')}{\sqrt{1+C'^2}\,\sqrt{m'^2+n'^2}} \,, \label{sin-beta_cos-beta}
\end{align}
so that
\begin{align}
    \tan \beta_0=\dfrac{n'-C'\,m'}{m'+C'\,n'}\,. \label{beta_cond}
\end{align}

Notice that the metric functions $\Omega, P, Q, $ in  \eqref{metr_shift} \emph{do not contain} $\beta$. As we shall see, this is of a great help in identifying the horizons and poles of these black hole spacetimes. The parameter $\beta$ appears only in the metric function $\rho$, defined in \eqref{rho-shift}, via \eqref{p0q0-in-ABR}. In particular, it determines the structure of the curvature singularity located at ${\rho^2=0}$.

More importantly, $\beta$ which enters $p_0$ and $q_0$ plays a crucial role in the character of the electromagnetic field described by the components \eqref{phi_0_shift}, \eqref{phi_1_shift}. In fact, in subsequent sections of this paper we will show that these parameters are related to \emph{charges of the black hole}.

\vspace{2mm}

It can immediately be seen from \eqref{defAB} that for ${C'=0}$ (keeping other parameters fixed) the constants $A', B', R'$ all vanish, and thus using \eqref{p0q0-in-ABR} we get ${p_0=0=q_0}$. Moreover, the conformal factor \eqref{Omega-C'} reduces to a very simple expression ${\Omega^2 = (1-pq)^2}$ (notice from \eqref{C' definition} that ${C'=0}$ necessarily implies ${c'=0}$, unless in a peculiar subcase ${k'=0}$), so that
\begin{align}
    \rho^2 = p^2+q^2\,,\qquad  \Omega = 1-pq\,.  \label{roho-Omega-PD}
\end{align}
The coefficients \eqref{a'-i}, \eqref{b'-i} also simplify considerably to
\begin{align}
    a'_0=b'_0=k'=-a'_4=-b'_4\,,\quad
    a'_1=-b'_3=2n'\,,\quad
   -a'_2=b'_2=\epsilon'\,,\quad
    a'_3=-b'_1=2m'\,,
\end{align}
and thus the metric functions \eqref{P-PD}, \eqref{Q-PD} reduce to
\begin{align}
   P&= k'+2n'p-\epsilon'p^2+2m'p^3-k'p^4\,, \label{P-PD0}\\
   Q&= k'-2m'q+\epsilon'q^2-2n'q^3-k'q^4\,. \label{Q-PD0}
\end{align}
For ${C'=0}$, that is for a vanishing electromagnetic field, we thus obtain the metric \eqref{metr},
\begin{align}
    \dd s^2=\frac{1}{\Omega^2}\Big[-\dfrac{Q}{\rho^2}(\dd \eta-p^2\dd\sigma)^2+\dfrac{\rho^2}{Q}\dd q^2
    +\dfrac{\rho^2}{P}\dd p^2+\dfrac{P}{\rho^2}(\dd \eta+q^2\dd \sigma)^2\Big], \label{PD}
\end{align}
which is exactly the class of \emph{type~D vacuum Pleba\'nski-Demia\'nski solutions}, see \cite{Plebanski1976} or Eqs.~(16.1), (16.2) in \cite{GriffithsPodolsky:2009}. This justifies the introduction of this parametrization, and calling it the Pleba\'nski-Demia\'nski-type form of the metric.

\section{Introducing the acceleration and twist parameters $\alpha$ and $\omega$}

The new class of type D solutions to Einstein-Maxwell equations we found in the previous Section is very large, depending of 5~real parameters $k', \epsilon', m', n', \beta$, and 1~complex $c'$. However, its physical interpretation is not  obvious, standard black holes are not identified, and its  form does not directly admit various geometrically distinct subcases. In particular, in the current metric (\ref{PD}) it is not even possible to obtain static subcases because the spacetime is always twisting (the spin coefficients $\rho_{\rm sc}$ and $\mu$ given by \eqref{opt_sc_5} always have non-zero imaginary part).

Actually, the Pleba\'nski-Demia\'nski (PD) solution \cite{Plebanski1976} in its original coordinates presented in 1976 suffered from the same problem. This was the main motivation for the series of works \cite{Griffiths2005,GRIFFITHS2006,PodolskyGriffiths:2006}  by Griffiths and Podolsk\'y (GP) in  2005--6, summarized in Chapter~16 of \cite{GriffithsPodolsky:2009}, in which \emph{two convenient parameters} with clear kinematic meaning were introduced, namely the \emph{acceleration} parameter~$\alpha$ and the \emph{twist} parameter~$\omega$. Indeed, for ${\alpha=0}$ the corresponding black holes do not accelerate, and for ${\omega=0}$ they do not rotate (because the imaginary part of ${\rho_{\rm sc}=\mu}$ vanishes).

The new metric (\ref{metr_shift}) resembles the Pleba\'nski-Demia\'nski metric \eqref{roho-Omega-PD}--\eqref{PD}, and reduces to it for ${C'=0}$. Therefore, we will apply the same trick as in \cite{GRIFFITHS2006}, that is, we will perform the specific rescaling (and trivial renaming) of the coordinates\footnote{In our recent studies \cite{Ovcharenko2024,Ovcharenko2025} we denoted this change as a transition from the PD metric to the PD$_{\alpha\omega}$ metric. It should also be emphasized that $\alpha$ here is the GP acceleration parameter, which is \emph{generally different} from the A$^+$ acceleration parameter, as investigated in full detail also in \cite{Ovcharenko2024,Ovcharenko2025}.}
\begin{equation}
     p=\sqrt{\alpha \omega}\,x\,, \qquad
     q=\sqrt{\dfrac{\alpha}{\omega}}\,r\,,\qquad
  \eta=\sqrt{\dfrac{\omega}{\alpha}}\,\tau\,,\qquad
\sigma=\sqrt{\dfrac{\omega}{\alpha^3}}\,\phi\,,
\end{equation}
accompanied by the rescaling of the parameters
\begin{align}
    &m'+\im\, n'=\sqrt{\dfrac{\alpha^3}{\omega^3}}\,(m+\im\,n)\,,\qquad
  \epsilon'=\dfrac{\alpha}{\omega}\,\epsilon\,,\qquad
    k'=\alpha^2k\,,  \nonumber\\[1mm]
    &c'=\sqrt{\dfrac{\omega}{\alpha}}\,c\,,\qquad
    C' = \alpha \omega\, C\,,\qquad
    p_0=\sqrt{\alpha\omega}\,x_0\,,\qquad
    q_0=\sqrt{\dfrac{\alpha}{\omega}}\,r_0\,.\label{rescl_params}
\end{align}
Note that these are exactly the transformations given by Eqs.~(3),~(4) in \cite{GRIFFITHS2006} (except that the complex electromagnetic charge $c'$ scales differently than in the PD case with just the aligned Maxwell field). Introducing the rescaled metric functions and coefficients as
\begin{align}
 &\mathcal{P}(x):=\dfrac{1}{\alpha^2}\,P\big(\sqrt{\alpha\omega}\,x\big),\qquad    \mathcal{Q}(r):=\dfrac{\omega^2}{\alpha^2}\,Q\Big(\sqrt{\dfrac{\alpha}{\omega}}\,r\Big),\qquad
 {\varrho}^2(r,x) :=\dfrac{\omega}{\alpha}\,\rho^2, \nonumber\\[2mm]
 &a_i :=\dfrac{(\alpha\omega)^{i/2}}{\alpha^2}\,a'_i\,,\qquad
  b_i :=\left(\dfrac{\omega}{\alpha}\right)^{(4-i)/2}\,b'_i \,,\label{a_i,b_i_def}
\end{align}
the metric  \eqref{metr_shift} becomes
\begin{align}
    \dd s^2 =\dfrac{1}{\Omega^2}\Big[&-\dfrac{\mathcal{Q}}{\varrho^2}\big[(\dd\tau-\omega\,(x+x_0)^2 \dd\phi\big]^2
    +\dfrac{\varrho^2}{\mathcal{Q}}\,\dd r^2\nonumber\\
   &+\dfrac{\mathcal{P}}{\varrho^2}\big[\,\omega\,\dd\tau+(r+r_0)^2\dd\phi\big]^2
    +\dfrac{\varrho^2}{\mathcal{P}}\,\dd x^2 \,\Big]\,,\label{ds_alpha_omega}
\end{align}
where the function $\varrho^2(r,x)$ and the conformal factor $\Omega^2(r,x)$ are given by
\begin{align}
  \varrho^2 &=(r+r_0)^2+\omega^2(x+x_0)^2\,,       \label{rho2_rescl}  \\
  \Omega^2  &= (1-\alpha rx)^2 +2\alpha^2C(r^2-\omega^2x^2)+\alpha^2 \Big(4\alpha^2\omega^2C^2-2C\dfrac{\epsilon}{k}\Big)\,r^2x^2 \nonumber\\
         & \hspace{18mm}+\dfrac{4\alpha^2C}{(1+\alpha^2\omega^2C^2)k}
         \Big[(m+\alpha \omega\, C n)\,x + (n-\alpha \omega\, Cm)\dfrac{r}{\omega}\,\Big]\,rx\,, \label{Om2_rescl}
\end{align}
and $\mathcal{P}, \mathcal{Q}$ are quartic polynomials
\begin{align}
    \mathcal{P}(x)&= a_0+a_1\,x+a_2\,x^2+a_3\,x^3+a_4\,x^4\,,\label{P_prime}\\
    \mathcal{Q}(r)&= b_0+b_1\,r+b_2\,r^2+b_3\,r^3+b_4\,r^4\,.\label{Q_prime}
\end{align}
Here the real constant is
\begin{equation}
  C := 2c\bar{c} k \,,
    \label{C definition}
\end{equation}
and the coefficients $a_i$ and $b_i$ defined in (\ref{a_i,b_i_def}) take the explicit form
\begin{align}
    a_0&= k\,, \nonumber\\
    a_1&= \frac{2}{\omega}\,\dfrac{n-\alpha\omega Cm}{1+\alpha^2\omega^2C^2}\,, \nonumber\\
    a_2&=-\epsilon \,,\label{prim_param_6}\\
    a_3&= 2\alpha\,\dfrac{(1+2\alpha^2\omega^2C^2)m-\alpha\omega Cn}{1+\alpha^2\omega^2C^2}\,, \nonumber\\
    a_4&=-\alpha^2\omega^2(1+4\alpha^2\omega^2C^2)k + 2\alpha^2\omega^2C \Big(\epsilon-\dfrac{(m+\alpha\omega Cn)^2}{\omega^2k(1+\alpha^2\omega^2C^2)^2}\Big)\,, \nonumber
\end{align}
and
\begin{align}
    b_0&= \omega^2 k\,, \nonumber\\
    b_1&=-2\,\dfrac{m+\alpha\omega Cn}{1+\alpha^2\omega^2C^2}\,, \nonumber\\
    b_2&= \epsilon \,,\label{prim_param_1}\\
    b_3&=-2\,\frac{\alpha}{\omega}\,\dfrac{(1+2\alpha^2\omega^2C^2)n+\alpha \omega C m}{1+\alpha^2\omega^2C^2}\,, \nonumber\\
    b_4&=-\alpha^2(1+4\alpha^2\omega^2C^2)k + 2\alpha^2C\Big(\epsilon+\dfrac{(n-\alpha\omega Cm)^2}{\omega^2k(1+\alpha^2\omega^2C^2)^2}\Big)\,. \nonumber
\end{align}
Finally, the constraint (\ref{p0q0_cond}) now becomes
\begin{align}
  (\omega\,x_0-A)^2  + (\,r_0-B)^2 = R^2\,,\label{r0x0_cond}
\end{align}
where the constants $A, B, R$ read
\begin{align}
A &= \dfrac{2C}
{(1+\alpha^2\omega^2C^2)[(1+4\alpha^2\omega^2C^2)k - 2C\epsilon]}\,(n-\alpha\omega\,C m)\,,\nonumber\\
B &= \dfrac{2C}
{(1+\alpha^2\omega^2C^2)[(1+4\alpha^2\omega^2C^2)k - 2C\epsilon]}\,(m+\alpha\omega\,C n)\,,\label{defABnew}\\
R &= \dfrac{2C}
{\sqrt{1+\alpha^2\omega^2C^2}\,[(1+4\alpha^2\omega^2C^2)k - 2C\epsilon]}\,\sqrt{m^2+n^2}\,.\nonumber
\end{align}
The relation \eqref{r0x0_cond} is satisfied by considering a single angular parameter $\beta$ such that
\begin{align}
 \omega\,x_0 &= A + R \sin\beta\,,\nonumber\\
         r_0 &= B + R\cos\beta\,.\label{x0r0-in-ABR}
\end{align}
The two coefficients $\omega\,x_0$ and $r_0$ can \emph{always be made zero} (${\omega\,x_0=0=r_0}$) by the special choice \eqref{beta_cond} of $\beta$, which now reads
\begin{align}
    \tan \beta_0=\dfrac{n\,-\alpha\omega\,m\,C}{m+\alpha\omega\,n\,C}\,. \label{beta_cond-new}
\end{align}

\vspace{2mm}

The components of the electromagnetic field \eqref{phi_0_shift} and \eqref{phi_1_shift} with respect to the null tetrad
\begin{align}
    \mathbf{k}&=\dfrac{\Omega}{\sqrt{2\mathcal{Q}}\,\varrho}\,
       (r^2\partial_{\tau}-\omega\,\partial_{\phi}+\mathcal{Q}\,\partial_{r}),\nonumber\\
    \mathbf{l}&=\dfrac{\Omega}{\sqrt{2\mathcal{Q}}\,\varrho}\,
       (r^2\partial_{\tau}-\omega\,\partial_{\phi}-\mathcal{Q}\,\partial_{r}),\nonumber\\
    \mathbf{m}&=\dfrac{\Omega}{\sqrt{2\mathcal{P}}\,\varrho}\,
       (\omega\,x^2\partial_{\tau}+\partial_{\phi}+\im\,\mathcal{P}\,\partial_{x}),\nonumber\\
    \bar{\mathbf{m}}&=\dfrac{\Omega}{\sqrt{2\mathcal{P}}\,\varrho}\,
       (\omega\,x^2\partial_{\tau}+\partial_{\phi}-\im\,\mathcal{P}\,\partial_{x}),\label{null_tetr_trans}
\end{align}
(which is obtained by the rescaling of (\ref{null_tetr})) are
\begin{align}
\Phi_0&=\Phi_2=\dfrac{\alpha c}{\Omega}\,\dfrac{\sqrt{\mathcal{P}\mathcal{Q}}}{(r+r_0)+\im\,\omega\,(x+x_0)}\,,
    \label{phi_0_rescaled} \\
\Phi_1&=\dfrac{1}{4\alpha\,\bar{c}}\,\dfrac{\Omega^2}{[(r+r_0)+\im\,\omega\,(x+x_0)]^2}\,\Big[\,
     \omega\,(x+x_0)^2\Big(\dfrac{\Omega_{,r}}{x+x_0}\Big)_{,x}
     -\im\,  (r+r_0)^2\Big(\dfrac{\Omega_{,x}}{r+r_0}\Big)_{,r} \,\Big].
     \label{phi_1_rescaled}
\end{align}
The corresponding  potential \eqref{A-2} has been rescaled to
\begin{align}
    \mathbf{A} = \dfrac{1}{4\alpha\,\bar{c}}\Big[\,
    \dfrac{ \omega\,\Omega_{,r}-\im\,\Omega_{,x}}{(r+r_0)+\im\,\omega\,(x+x_0)}\,\dd \tau
    +\Big(\dfrac{(r+r_0)^2\,\Omega_{,r}+\im\,\omega\,(x+x_0)^2\,\Omega_{,x}}{(r+r_0)+\im\,\omega\,(x+x_0)}-\Omega\Big)\,\dd\phi
    \Big].   \label{A-3}
\end{align}

\vspace{4mm}

In the reparametrized metric \eqref{ds_alpha_omega} it is now easy to obtain the \emph{static} (non-twisting) subcase by simply setting ${\omega=0}$, in which case the metric becomes \emph{diagonal}.

Taking the acceleration $\alpha$ to zero is not so straightforward, and depends on how do other parameters behave in this limit.

If we set the acceleration to zero by ${\alpha=0}$, while \textit{keeping} $|c|$ \textit{constant}, then the conformal factor simplifies enormously to ${\Omega=1}$, and the metric functions are just
\begin{align}
    \mathcal{P} = k+2n/\omega\,x-\epsilon \,x^2\,,\qquad
    \mathcal{Q} = \omega^2 k-2m\,r+\epsilon\,r^2\,.\label{PQ_alpha=0}
\end{align}
Even more interestingly, in this limit \eqref{phi_0_rescaled} become
\begin{align}
\Phi_0=0=\Phi_2\,.\label{phi0,2_alpha=0}
\end{align}
This means that \emph{for such non-accelerating black holes the non-aligned components $\Phi_0$ and $\Phi_2$ of the electromagnetic field vanish}. As we will show, in such a case, we recover the Kerr-Newman-NUT black holes.

Interestingly, this is not the only acceleration-free limit. As we will show further in Sec.~\ref{subsection e=0=g and l=0}, when both $r_0$ and $x_0$ are zero it is possible to perform the $\alpha\rightarrow 0$ limit while $|c|\rightarrow \infty$ \textit{such that} $2\alpha|c|\equiv B=\mathrm{const}$. This special limit represents the \textit{Kerr black hole in a uniform magnetic field} of the Bertotti-Robinson type, presented by us in \cite{Kerr-BR}.

However, before we describe how such limits can be taken, we have to investigate the structure of the electromagnetic field in more detail, which is done in the next Section.

\section{The electromagnetic field}
\label{sec_em_field}

The non-trivial electromagnetic field in the new class of solutions presented here is the main ingredient that distinguishes these spacetimes from other (well-known) black holes of algebraic type D, for which the electromagnetic field is aligned with the geometry. Focusing on its properties is thus key for the interpretation of these solutions, and understanding the physical role of its parameters.

The acceleration and twist parameters $\alpha$ and $\omega$ allowed us to put the expressions \eqref{phi_0_shift}, \eqref{phi_1_shift} for the electromagnetic scalars $\Phi_0$, $\Phi_1$, $\Phi_2$ into the form \eqref{phi_0_rescaled}, \eqref{phi_1_rescaled}.

We can thus easily obtain the \emph{static (non-twisting)} limit by simply setting ${\omega=0}$, so that
\begin{align}
\Phi_0& =\Phi_2=\,\dfrac{\alpha c}{\Omega}\,\dfrac{\sqrt{\mathcal{P}\mathcal{Q}}}{r+r_0}\,,\label{phi_02_omega=0}\\
\Phi_1&
=-\dfrac{\im}{4}\dfrac{\Omega^2}{\alpha\,\bar{c}}\,(r+r_0)
\Big(\dfrac{\Omega_{,x}}{r+r_0}\Big)_{,r}\,. \label{phi_1_omega=0}
\end{align}
The expression for ${\Phi_0=\Phi_2}$ exactly corresponds to the results of Van den Bergh and Carminati presented in Eq.~(14) of \cite{VandenBergh2020}.

The \emph{non-accelerating} limit ${\alpha=0}$, as we already mentioned, is not unique. Before considering it, let us simplify the expression for $\Phi_1$ given by \eqref{phi_1_rescaled}, which can be rewritten as
\begin{align}
    \Phi_1&=\dfrac{1}{4\alpha\,\bar{c}}\dfrac{\Omega^2}{[(r+r_0)+\im\,\omega (x+x_0)]^2}\,\Big[\,
     \omega\,x^2\Big(\dfrac{\Omega_{,r}}{x}\Big)_{,x}
     -\im\,  r^2\Big(\dfrac{\Omega_{,x}}{r}\Big)_{,r}
     +(\omega\,x_0-\im\,r_0)\,\Omega_{,rx}\Big].
\end{align}

Now let us express the complex constant $c$ in the \emph{polar form}
\begin{align}
    c \equiv |c|\,{\rm e}^{\im\,\gamma}\,, \label{def_gamma}
\end{align}
and define two new quantities as \emph{charges}
\begin{align}
    e :=-\dfrac{1}{2|c|}\,\big(\omega\,x_0\,\cos\gamma + r_0\,\sin\gamma\big)\,,\qquad
    g :=-\dfrac{1}{2|c|}\,\big(\omega\,x_0\,\sin\gamma - r_0\,\cos\gamma\big)\,, \label{eg_def}
\end{align}
or inversely,
\begin{align}
       r_0 = -2|c|\,\big(e\cos\gamma - g\sin\gamma \big),\qquad
\omega\,x_0 = -2|c|\,\big(e\sin\gamma + g\cos\gamma \big). \label{r0x0-given-by-eg}
\end{align}
Consequently, the charges satisfy the relation
\begin{align}
    e^2+g^2=\dfrac{1}{4|c|^2}\big(r_0^2+\omega^2 x_0^2\big)\,. \label{e2+g2=r2+omega2}
\end{align}
Recall also the expressions \eqref{x0r0-in-ABR}, namely ${\omega\,x_0 = A + R \sin\beta}$ and ${r_0 = B + R\cos\beta}$, so that
\begin{align}
    e &= -\dfrac{1}{2|c|}\,\Big( (A\cos\gamma + B\sin\gamma )
         + R\,\sin(\beta+\gamma)\Big),\nonumber\\
    g &= -\dfrac{1}{2|c|}\,\Big( (A\sin\gamma - B\cos\gamma )
         - R\,\cos(\beta+\gamma)\Big), \label{eg_another-form}
\end{align}
and substituting into \eqref{r0x0-given-by-eg} we get
\begin{align}
       r_0 &= (A\cos2\gamma + B\sin2\gamma )
         + R\,\sin(2\gamma+\beta)\,,\nonumber\\[2mm]
\omega\,x_0 &= (A\sin2\gamma - B\cos2\gamma )
         - R\,\cos(2\gamma+\beta)\,. \label{r0-x0_another-form}
\end{align}
The constants $A, B, R$ are explicitly given by \eqref{defABnew}. They depend on $C$ defined in \eqref{C definition}
as ${C = 2 |c|^2 k}$, which means that \emph{they are independent of the angles $\beta$ and $\gamma$}. This enables us to interpret \eqref{eg_another-form} and \eqref{r0-x0_another-form} as specific forms of \emph{duality rotation represented by the two angular parameters $\beta$ and $\gamma$}. They ``mix'' the charges $e,g$ and also the related parameters $r_0, \omega x_0$. In particular, for any fixed ${\gamma=\gamma_0}$ (that is the complex phase of $c$), the value of the ``electric charge~$e$'' is effectively determined by $\,\sin(\beta+\gamma_0)$, while the value of the ``magnetic charge~$g$'' is determined by $-\cos(\beta+\gamma_0)$. Recall also that for the special choice ${\beta=\beta_0}$ given by \eqref{beta_cond-new} the coefficients $\omega\,x_0$ and $r_0$ are \emph{both made zero} (${\omega\,x_0=0=r_0}$). The meaning of the parameter $\gamma$ is not fully clear now, as it represents the duality rotation of both aligned $\Phi_1$ and non-aligned $\Phi_0,~\Phi_2$ components of the electromagnetic field. We will give it some interpretation further.
\vspace{2mm}

We can thus see that the \emph{electromagnetic field is determined by 3 real parameters}, namely $|c|$ and two angular parameters $\beta, \gamma$ representing specific duality rotations described above.
In terms of these parameters the component $\Phi_1$ takes the form
\begin{align}
\hspace{-3mm}
    \Phi_1=\dfrac{1}{4\alpha}\,\dfrac{\Omega^2}{[(r+r_0)+\im\,\omega (x+x_0)]^2}\Bigg[
    \dfrac{{\rm e}^{\im\,\gamma}}{|c|} \Big[\,
    \omega\,x^2\Big(\dfrac{\Omega_{,r}}{x}\Big)_{,x}
     -\im\,r^2 \Big(\dfrac{\Omega_{,x}}{r}\Big)_{,r}\,\Big]
     -2(e+\im\,g)\,\Omega_{,rx}\Bigg].\label{phi_1_rescl_2}
\end{align}

Now, in the ${\alpha\rightarrow 0}$ limit of \emph{vanishing acceleration} (while keeping $|c|=\mathrm{const.}$), the conformal factor (\ref{Om2_rescl}) reduces to
${\Omega^2 = 1-2\alpha rx + O(\alpha^2)}$, so that ${\Omega = 1-\alpha rx + O(\alpha^2)}$. The aligned $\Phi_1$ component of the electromagnetic thus considerably simplifies to
\begin{align}
    \Phi_1=\dfrac{\tfrac{1}{2} (e+\im\,g)}{[(r+r_0)+\im\,\omega (x+x_0)]^2}\,.\label{phi_1_for_alpha=0}
\end{align}
This expression agrees with the result for Kerr-Newman-NUT black holes with fully aligned (Coulombic) electromagnetic field (see e.g. \cite{GriffithsPodolsky:2009, Podolsk2023}), in which case the two parameters $e$~and~$g$ have direct physical interpretation as electric and magnetic charges, respectively. This means that in such a limit $\alpha\to 0$ we can interpret the parameters $e$ and $g$ (and equivalently the constants $r_0$ and $x_0$) as \emph{related to electric and magnetic charges of a black hole}.

However, it does not mean that these parameters \textit{are equal} to electric and magnetic charges of a black hole \textit{in the most general case} $\alpha\neq 0$. The corresponding physical charges may be found by integrating the electric and magnetic flux through the horizon of a black hole, but this exceeds the scope of our present work.


\section{The Griffiths-Podolsk\'y-type form}\label{Section_GP_form}

In the preceding Sections we have found a new class of spacetimes with the non-aligned electromagnetic field, generalizing the known type~D black hole
solutions. In fact, this solution is more general also in the context of the possible horizon topology, as currently this is not specified. For realistic black holes we expect the topology of their horizons to be spherical, and thus it is desirable to transform the coordinates and reparametrize the general solution so that we can restrict to the subclass of spherical topologies.

The same problem existed for the Pleba\'nski-Demia\'nski spacetime. The corresponding transformations were introduced by Griffiths and Podolsk\'y in \cite{Griffiths2005,GRIFFITHS2006,PodolskyGriffiths:2006}, where they obtained the Boyer-Lindquist-type coordinates for the PD class. We can apply the same method here, using the convenient fact that the metric \eqref{ds_alpha_omega} is very similar to the Pleba\'nski-Demia\'nski metric \eqref{PD}, with quartic functions $P(x)$ and $Q(r)$ given by \eqref{P-PD0}, \eqref{Q-PD0}.

Specifically, we will conduct a linear transformation of coordinates
\begin{align}
    x=\dfrac{a}{\omega}\,\tilde{x}+\dfrac{l}{\omega}\,,\qquad
 \tau=t-\dfrac{(a+l+\omega\,x_0)^2}{a}\,\varphi\,,\qquad
 \phi=-\dfrac{\omega}{a}\,\varphi\,, \label{GP_trans}
\end{align}
with the corresponding redefinition
\begin{align}
    \tilde{P}(\tilde{x}):=\dfrac{\omega^2}{a^2}\,\mathcal{P}\Big(\,\dfrac{a}{\omega}\,\tilde{x}+\dfrac{l}{\omega}\Big),
\end{align}
that puts the metric \eqref{ds_alpha_omega} into
\begin{align}
    \dd s^2=\dfrac{1}{\Omega^2}\Bigg[&-\dfrac{\mathcal{Q}}{\varrho^2}
    \Big(\dd t-\big[a(1-\tilde{x}^2)+2(l+\omega\, x_0)(1-\tilde{x})\big]\dd\varphi\Big)^2
    +\dfrac{\varrho^2}{\mathcal{Q}}\,\dd r^2 \nonumber\\
    &+\dfrac{\tilde{P}}{\varrho^2}\Big(a\,\dd t-\big[(r+r_0)^2+(a+l+\omega\,x_0)^2\big]\dd\varphi\Big)^2
    +\dfrac{\varrho^2}{\tilde{P}}\,\dd\tilde{x}^2\Bigg], \label{metr_GP_1}
\end{align}
where
\begin{align}
    \varrho^2=(r+r_0)^2+(a\,\tilde{x}+l+\omega\,x_0)^2.
\end{align}
The function $\tilde{P}(\tilde{x})$ has the quartic form
\begin{align}
    \tilde{P}=\tilde{a}_0+\tilde{a}_1\,\tilde{x}+\tilde{a}_2\,x^2+\tilde{a}_3\,\tilde{x}^3+\tilde{a}_4\,\tilde{x}^4\,,\label{tild_P}
\end{align}
where
\begin{align}
    \tilde{a}_0&= \dfrac{1}{a^2\omega^2}\,(\omega^4 a_0+l\omega^3 a_1+l^2\omega^2 a_2+l^3\omega\, a_3+l^4 a_4)\,,\nonumber\\
    \tilde{a}_1&= \dfrac{1}{a\,\omega^2}\,(\omega^3a_1+2l\omega^2 a_2+3l^2\omega\, a_3+4l^3 a_4)\,,\nonumber\\
    \tilde{a}_2&= \dfrac{1}{\omega^2}\,(\omega^2 a_2+3l \omega\,a_3+6 l^2a_4)\,, \label{tilda_ai}\\
    \tilde{a}_3&= \dfrac{a}{\omega^2}\,(\omega\,a_3+4l\,a_4)\,,\nonumber\\
    \tilde{a}_4&= \dfrac{a^2}{\omega^2}\,a_4\,,\nonumber
\end{align}
while $\mathcal{Q}(r)$ remains to be given by \eqref{Q_prime}.

The coefficients $\tilde{a}_i$ of $\tilde{P}$, explicitly expressed in terms of $k, m, n,\epsilon, C, \alpha, \omega$ using \eqref{prim_param_6}, are very complicated. However, as was argued in \cite{Griffiths2005,GRIFFITHS2006}, looking for the \emph{black hole solutions with spherical topology} the function $\tilde{P}$ \textit{must have at least two roots} (corresponding to two poles along the axis of symmetry). In this case  it is always possible to choose the parameters $a$ and $l$ in such a way that these roots are conveniently placed at ${\tilde{x}=1}$ and ${\tilde{x}=-1}$. (It will turn out that $a$ is the Kerr-like rotational parameter, $l$ in the NUT-like twist parameter, and ${\tilde{x}=\pm 1}$ will correspond to the poles at ${\theta=0}$ and ${\theta=\pi}$ in Boyer-Lindquist-type coordinates.) Therefore, the metric function $\tilde{P}$ may be written in the factorized form
\begin{align}
    \tilde{P}=(1-\tilde{x}^2)(\tilde{a}_0-\tilde{a}_3\,\tilde{x}-\tilde{a}_4\,\tilde{x}^2)\,.
    \label{factorized-P}
\end{align}
This natural requirement is satisfied if the coefficients $\tilde{a}_i$ given by \eqref{tilda_ai} obey the following two constraints
\begin{align}
    \tilde{a}_1+\tilde{a}_3=0\,,\qquad \tilde{a}_0+\tilde{a}_2+\tilde{a}_4=0\,.
\end{align}
Interestingly, they are \emph{linear} in $\epsilon$. By equating $\epsilon$ obtained independently from both of them, we get a \emph{quadratic} equation for~$n$. It is thus possible to express these two coefficients $n$~and~$\epsilon$ explicitly in terms of the remaining physical parameters, in particular $a$ and $l$, as
\begin{align}
n= &\ \dfrac{k}{4\alpha^4C^3  (a^2-l^2)l}
   \Bigg[ - 4 \alpha^3C^2(a^2-l^2)\dfrac{ml}{\omega k}  \nonumber\\
   & \quad + \Big(1 - \alpha^2 C \big[3a^2+5l^2-2\alpha^2 C (a^2-l^2)^2\big]  \mp I_1 I_2 \Big)
   \big(1+\alpha^2C^2\omega^2\big)   \,   \Bigg],\label{n_rel}\\
\epsilon = &\ \dfrac{k}{4\alpha^4C^3 (a^2-l^2)^2l^2}
    \Bigg[4\alpha^3 C^2\Big(a^4+2a^2l^2-3l^4-2\alpha^2C(a^2-l^2)^3\Big)\dfrac{ml}{\omega k}
    +8 \alpha^4 C^3\omega^2(a^2+l^2)l^2  \nonumber\\
   & \quad - \Big(1-\alpha^2 C\big[3a^2+5l^2-2\alpha^2C(a^2-l^2)^2\big] \mp I_1 I_2 \Big)
      \big(a^2+3l^2-\alpha^2 C(a^2-l^2)^2\big) \Bigg],\label{eps_rel}
\end{align}
where the new constants $I_1, I_2$ are convenient shorthands for the square roots
\begin{align}
I_1& := \sqrt{1 - \alpha^2 C \Big[2a^2+6l^2 - \alpha^2C (a^2-l^2)^2 + 8\alpha^2C^2 \omega^2 l^2
      + 8 \alpha C (a^2-l^2)\,\frac{ml}{\omega k}\,\Big]} \,,\label{s1-and-s2}\\
I_2& := \sqrt{1-4\alpha^2 C \big[a^2+l^2-\alpha^2C(a^2-l^2)^2\big]}\,. \nonumber
\end{align}

Equations (\ref{n_rel}) and (\ref{eps_rel}) explicitly relate the PD parameters $n$ and $\epsilon$ to the GP parameters $a$ and $l$. Finally, we can substitute them into the expression \eqref{tilda_ai} for~$\tilde{a}_0$ which yields
\begin{align}
\tilde{a}_0=\dfrac{k}{4\alpha^4 C^3(a^2-l^2)^2}
    & \Bigg[4\alpha^4\omega^2C^3(a^2-4l^2)
     -4\alpha^3C^2(a^2-l^2) \big[3-2\alpha^2C(a^2-l^2)\big] \frac{ml}{\omega k}  \label{b0tild}\\
    & +\big[3-\alpha^2C(a^2-l^2)\big]\Big(1-\alpha^2C\big[3a^2+5l^2-2\alpha^2C(a^2-l^2)^2\big]\mp I_1 I_2\Big)\Bigg].\nonumber
\end{align}
Recall that the real dimensionless constant $C$ was introduced in \eqref{C definition} as
\begin{equation}
C = 2k\, |c|^2\,, \label{C def-again}
\end{equation}
where the \emph{complex} parameter $c$ is related to the value of the \emph{non-aligned} components ${\Phi_0=\Phi_2}$ of the electromagnetic field \eqref{phi_0_rescaled}.

\vspace{2mm}

The character of the spacetime depends on whether $\tilde{a}_0$ is zero, positive, or negative. If it is non-zero, the scaling freedom can be used to set ${\tilde{a}_0=\pm 1}$. Effectively, this also determines the geometry of the \emph{black hole horizons} (located at ${\mathcal{Q}=0}$) with
\begin{align}
\tilde{a}_0=1  \label{a0=1}
\end{align}
corresponding to the usual \emph{compact, spherical-like topology}. Thus naturally assuming ${\tilde{a}_0=1}$, the equation \eqref {b0tild} determines the PD parameter $k$. By substituting the expressions for ${I_1, I_2}$ (which are square roots) it can be infered that it is the 5-th order polynomial equation, so that \textit{in general it has up to 5 roots}. This may be surprising, because in the case of Pleba\'nski-Demia\'nski spacetimes with aligned electromagnetic field it was shown in \cite{Griffiths2005, GRIFFITHS2006} that for given parameters $m,~a,~l,~\alpha$ there is \emph{only one unique solution} for ${k,~n,~\epsilon}$, namely
\begin{align}
k &= \dfrac{1 +2\alpha\,\dfrac{l}{\omega}\, m  -3\alpha^2\dfrac{l^2}{\omega^2}( e^2+g^2 )}{\dfrac{\omega^2}{a^2-l^2} + 3\alpha^2\,l^2}\,, \label{k_GP} \\[2mm]
  n &= \frac{\omega^2k}{a^2-l^2}\,l -\alpha\,\frac{a^2-l^2}{\omega}\, m
 +\alpha^2\,l\, \frac{a^2-l^2}{\omega^2}\,(\omega^2k+ e^2+g^2 ),
  \label{n_GP}\\[5mm]
\epsilon &= \frac{\omega^2k}{a^2-l^2} +4\alpha\,\frac{l}{\omega}\, m
 -\alpha^2\, \frac{a^2+3l^2}{\omega^2}\,(\omega^2k+ e^2+g^2 ).
  \label{eps_GP}
\end{align}
In the present more general case, we have up to 5 roots. However, there is actually no contradiction because in the limit ${|c|\to 0}$, four of these roots diverge (or disappear).

\begin{figure}
    \centering
  \includegraphics[width=1\linewidth]{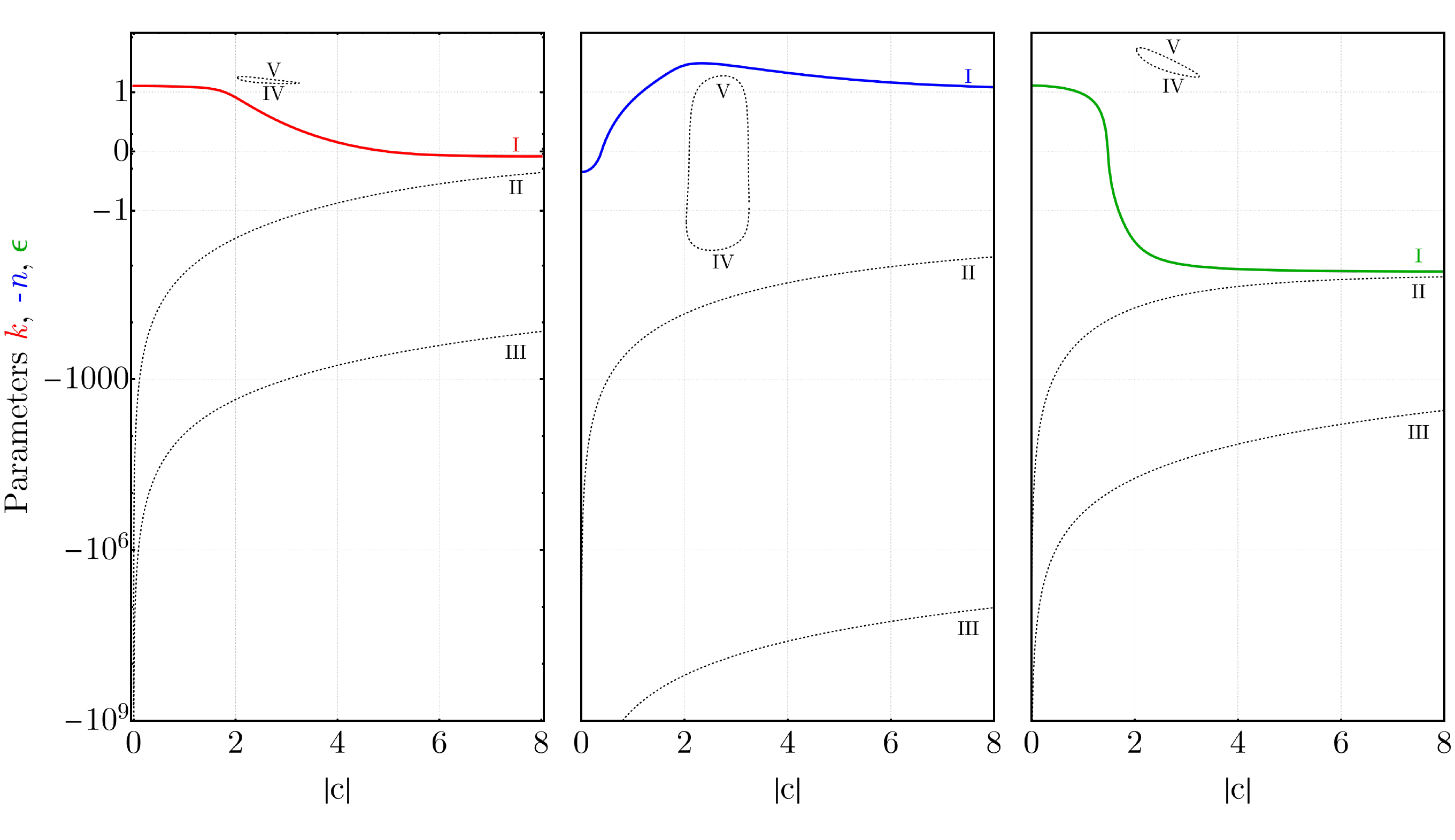}
    \caption{Graphs showing the dependence of the Pleba\'nski-Demia\'nski parameters $k$, $-n$, $\epsilon$ on $|c|$, calculated from \eqref{n_rel}, \eqref{eps_rel}, \eqref{b0tild}, plotted in symmetric logarithmic scale. The Roman numbers I-V label different 5 possible roots. Colored solid lines represent the solutions with \emph{finite} real limits as ${|c|\rightarrow 0}$ (these are the upper lines denoted as~I), while the dashed lines represent the roots of \eqref{n_rel}, \eqref{eps_rel}, \eqref{b0tild} which either diverge or are complex in the ${|c|\rightarrow 0}$ limit. Notice also that the roots IV and V occur only in a small restricted interval of ${|c|\ne0}$. The specific fixed values of the physical parameters are ${m=2.2}$, ${a=1.1}$, ${l=0.2}$, ${\omega=1}$, ${\alpha=0.14}$.}
    \label{kne_plot}
\end{figure}

This is nicely seen in Fig.~\ref{kne_plot} which plots all the admitted solutions for the PD parameters ${k, -n, \,\epsilon\,}$ as functions of the parameter $|c|$. It is clearly seen that \emph{only one solution remains finite} as ${|c|\to 0}$, denoted in Fig.~\ref{kne_plot} by the Roman number~I (the solid colored line). It corresponds to the upper (minus) sign in front of $I_1I_2$ in the expressions (\ref{n_rel}), (\ref{eps_rel}), (\ref{b0tild}) in the limit  ${C\to 0}$. Moreover, such a finite solution exactly agrees with the expressions \eqref{k_GP}--\eqref{eps_GP} for the Pleba\'nski-Demia\'nski black holes of algebraic type D \cite{Plebanski1976} in the Griffiths-Podolsk\'y representation (see Eqs.~(14)--(16) in \cite{GRIFFITHS2006}, and Eqs.~(16.15)--(16.17) in \cite{GriffithsPodolsky:2009}) in case of vacuum (${e=0=g}$, ${\Lambda=0}$). The role of the other four classes of solutions, denoted as II, III, IV, V, has to be understood better, but this exceeds the scope of the current work.

\newpage

Now, taking this finite solution of (\ref{n_rel}), (\ref{eps_rel}), and (\ref{b0tild}) with (\ref{a0=1}), the metric function~$\tilde{P}$ simplifies to
${\tilde{P}=(1-\tilde{x}^2)(1-\tilde{a}_3\,\tilde{x}-\tilde{a}_4\,\tilde{x}^2)}$. The final simple step is to introduce the spherical-like coordinate~$\theta$ by ${\tilde{x}=\cos\theta}$, so that the metric \eqref{metr_GP_1} takes form
\begin{align}
    \dd s^2=\dfrac{1}{\Omega^2}\Bigg[&-\dfrac{\mathcal{Q}}{\varrho^2}
    \Big(\dd t-\big[a\sin^2\theta+2(l+\omega\, x_0)(1-\cos\theta)\big]\dd\varphi\Big)^2
    +\dfrac{\varrho^2}{\mathcal{Q}}\,\dd r^2 \nonumber\\
    &+\dfrac{\varrho^2}{\tilde{\mathcal{P}}}\dd\theta^2+\dfrac{\tilde{\mathcal{P}}}{\varrho^2}\sin^2\theta
    \Big(a\,\dd t-\big[(r+r_0)^2+(a+l+\omega\, x_0)^2\big]\dd\varphi\Big)^2\Bigg],\label{GP_metr}
\end{align}
with the metric functions
\begin{align}
    \tilde{\mathcal{P}}(\theta) & =1-\tilde{a}_3\cos\theta-\tilde{a}_4\cos^2\theta\,, \label{tilde_P}\\
    \mathcal{Q}(r)      & = b_0+b_1\,r+b_2\,r^2+b_3\,r^3+b_4\,r^4\,,\label{math-Q}\\
    \varrho^2(r,\theta) & = (r+r_0)^2+(a\cos \theta+l+\omega\,x_0)^2\,,  \label{rho2_rescl_GP}
\end{align}
where the coefficients are explicitly given by (\ref{tilda_ai}), (\ref{prim_param_6}), (\ref{prim_param_1}), (\ref{x0r0-in-ABR}). The conformal factor $\Omega^2(r,\theta)$ is determined by (\ref{Om2_rescl}), (\ref{GP_trans}), that is
\begin{align}
  \Omega^2  &= \Big[1-\dfrac{\alpha}{\omega}(a\cos\theta+l)r\Big]^2 \nonumber\\
     & \hspace{4mm} +2\alpha^2C\big[r^2-(a\cos\theta+l)^2\big]
        +2\alpha^2C\Big(2\alpha^2C-\frac{\epsilon}{\omega^2k}\Big)(a\cos\theta+l)^2\,r^2 \label{Om2_rescl_GP}\\
     & \hspace{4mm} + \dfrac{4\alpha^2 C}{(1+\alpha^2\omega^2C^2)\omega^2 k}
         \Big[(m+\alpha \omega\, C n)(a\cos\theta+l) + (n-\alpha \omega\, Cm)\,r\,\Big](a\cos\theta+l)\,r\,. \nonumber
\end{align}

The non-aligned components (\ref{phi_0_rescaled}) of the electromagnetic field now read
\begin{align}
    \Phi_0=\Phi_2=\alpha\,c\,\dfrac{a}{\omega}\,
        \dfrac{\sqrt{\tilde{\mathcal{P}}\mathcal{Q}}}
        {(r+r_0)+\im\,(a\cos\theta+l+\omega \,x_0)}\,\dfrac{\sin\theta}{\Omega}\,,\label{phi_0-GP}
\end{align}
while the aligned component $\Phi_1$ is obtained from (\ref{phi_1_rescaled}) by the substitution ${\omega\,x=a\cos\theta+l}$,
\begin{align}
\Phi_1=\dfrac{1}{4\alpha\,\bar{c}\,\sin\theta}&\,\dfrac{\omega}{a}\,
    \dfrac{\Omega^2}{[(r+r_0)+\im\,(a\cos\theta+l+\omega \,x_0)]^2} \label{phi_1-GP}\\
    &\times\Big[
     (a\cos\theta+l+\omega \,x_0)^2\Big(\dfrac{-\Omega_{,r}}{a\cos\theta+l+\omega \,x_0}\Big)_{,\theta}
     +\im\,  (r+r_0)^2\Big(\dfrac{\Omega_{,\theta}}{r+r_0}\Big)_{,r} \,\Big].\nonumber
\end{align}
The corresponding  potential 1-form $\mathbf{A}$ follows from \eqref{A-3} by the same substitution, namely
\begin{align}
    \mathbf{A} = \dfrac{1}{4\alpha\,\bar{c}}\,\dfrac{\omega}{a}\Bigg[\,
    &\Omega_{,r}\,\dfrac{ a\,\dd t-[(r+r_0)^2+(a+l+\omega\, x_0)^2]\,\dd\varphi}
        {(r+r_0)+\im\,(a\cos\theta+l+\omega \,x_0)}         \label{A-4}\\
    +&\dfrac{\im\,\Omega_{,\theta}}{\sin\theta}\,\dfrac{\dd t
    -[a\sin^2\theta+2(l+\omega\, x_0)(1-\cos\theta)]\,\dd\varphi}
       {(r+r_0)+\im\,(a\cos\theta+l+\omega \,x_0)}  + \Omega\,\dd\varphi\, \Bigg]+\mathbf{A}_0\,.\nonumber
\end{align}
There is also a gauge freedom in choosing $\mathbf{A}_0$ such that ${\dd\mathbf{A}_0=0}$. For example, ${\mathbf{A}_0=-\tfrac{1}{4\alpha\,\bar{c}}\,\tfrac{\omega}{a}\,\dd\varphi}$ removes the apparent divergence of $\mathbf{A}$ as ${\alpha\to0},~|c|=\mathrm{const}$.

It is important to note what role the twist parameter $\omega$ plays in this solution. In \cite{Ovcharenko2024} we showed for the Pleba\'nski-Demia\'nski spacetime that, if at least one of the parameters $a$ and $l$ is non-zero, then $\omega$ can be chosen arbitrarily and it represents just the rescaling of the acceleration parameter $\alpha$. As our case is somewhat analogous, the interpretation of the parameter $\omega$ remains the same. The case when both $a$ and $l$ are zero is not considered in this work, and it is postponed to \cite{OP-prepar}.
\newpage

There also exists an \emph{alternative}, closely related compact form of the metric for this large class of black hole solutions. It is obtained by performing a simple shift of the coordinate $r$ and reparametrization of the NUT constant to $\tilde{l}$ such that
\begin{align}
    \tilde{r}:=r+r_0\,,\qquad
    \tilde{l}:=l+\omega\,x_0\,. \label{GP_trans_2}
\end{align}
The metric \eqref{GP_metr} thus simplifies to
\begin{align}
    \dd s^2=\dfrac{1}{\Omega^2}\Bigg[&-\dfrac{\mathcal{Q}}{\varrho^2}
    \Big(\dd t-\big[a\sin^2\theta
    +2\tilde{l}(1-\cos\theta)\big]\dd\varphi\Big)^2
    +\dfrac{\varrho^2}{\mathcal{Q}}\,\dd \tilde{r}^2 \nonumber\\
    &+\dfrac{\varrho^2}{\tilde{\mathcal{P}}}\,\dd\theta^2+\dfrac{\tilde{\mathcal{P}}}{\varrho^2}\sin^2\theta
    \Big(a\,\dd t-\big[\tilde{r}^{\,2}+(a+\tilde{l}\,)^2\big]\dd\varphi\Big)^2\Bigg],\label{GP_metr_2}
\end{align}
where
\begin{align}
    \varrho^2(\tilde{r},\theta)=\tilde{r}^2+(a\cos\theta+\tilde{l}\,)^2,
\end{align}
Such metric is exactly the Griffiths-Podolsk\'y general form of type D black holes with (aligned) Maxwell field, see Eq.~(16.18) in \cite{GriffithsPodolsky:2009}, but the 1-form potential is now more complicated and given by
\begin{align}
    \mathbf{A} = \dfrac{1}{4\alpha\,\bar{c}}\,\dfrac{\omega}{a}\Big[\,
    &\Omega_{,\tilde{r}}\,\dfrac{ a\,\dd t-\big[\,{\tilde{r}}^{\,2}+(a+\tilde{l}\,)^2\big]\dd\varphi }
        {\tilde{r}+\im\,(a\cos\theta+\tilde{l}\,)}
            \label{A-6}\\
    +&\dfrac{\im\,\Omega_{,\theta}}{\sin\theta}\,
     \dfrac{ \dd t - \big[a\sin^2\theta+2\tilde{l}(1-\cos\theta)\big]\dd\varphi}
       {\tilde{r}+\im\,(a\cos\theta+\tilde{l}\,)} + \Omega\,\dd\varphi\, \Big]  + \mathbf{A}_0\,.\nonumber
\end{align}
Notice that this explicit form of the electromagnetic field involves (in the numerators) exactly the 1-forms which appear in the metric \eqref{GP_metr_2}.


\vspace{5mm}

The constants $r_0$ and $x_0$, related to the charges $e$ and $g$ via \eqref{r0x0-given-by-eg}, are thus removed from the metric \eqref{GP_metr_2} and its function $\varrho^2$, and also from the potential \eqref{A-6}. However, they are contained \textit{implicitly} in the remaining metric functions $\tilde{\mathcal{P}}$, $\mathcal{Q}$ and $\Omega^2$ because it is necessary to replace the auxiliary parameter $l$ by ${\tilde{l}-\omega\,x_0}$. They also appear in the equations for $\epsilon,~n$ and $k$ (see (\ref{n_rel}), (\ref{eps_rel}) and (\ref{b0tild})) which additionally complicates the corresponding formulas. The expression for $\mathcal{Q}(\tilde{r})$ and for $\Omega^2(\tilde{r},\theta)$ become even more complicated by the substitution ${r=\tilde{r}-r_0}$ in the quartic \eqref{math-Q} and in \eqref{Om2_rescl_GP}, respectively. Because of these complications, we do not consider this form in the general case, only in some of the special subcases.

\newpage

\section{Important particular cases}

In previous Sections we derived the new class of solutions to the Einstein-Maxwell field equations of algebraic type D with a non-aligned electromagnetic field, and we presented it in various metric forms. This general solution is quite complicated, so it may be illustrative to consider now its particular cases. After understanding the main features of these cases, we will return to the investigation of the most general situation in the final Section~\ref{Sec-Physics}.

\subsection{No acceleration (${\alpha=0},~|c|=\mathrm{const.}$): Kerr-Newman-NUT}\label{gen_no_accel}

First, we will focus on spacetimes with zero acceleration parameter, ${\alpha=0}$ while keeping the complex parameter $c=\mathrm{const}$. The conformal factor (\ref{Om2_rescl_GP}) becomes simply ${\Omega=1}$, and the coefficients $a_i,~b_i$, given by (\ref{prim_param_6})--(\ref{prim_param_1}), reduce to
\begin{align}
    a_0&=k\,,\quad\qquad a_1=2\,\dfrac{n}{\omega}\,,\qquad a_2=-\epsilon\,,\qquad a_3=0,\qquad a_4=0\,,
       \label{a_i-for-alpha=0}\\
    b_0&=\omega^2k\,,\qquad b_1=-2m\,,\qquad b_2=\epsilon\,,\qquad b_3=0\,,\qquad b_4=0\,.
       \label{b_i-for-alpha=0}
\end{align}
The electromagnetic field \eqref{phi_0-GP}, \eqref{phi_1-GP} simplifies to
\begin{align}\label{Phi_2-for-alpha=0}
    \Phi_0 = 0 = \Phi_2 \,,\qquad
    \Phi_1 = \dfrac{\tfrac{1}{2}(e+\im\, g)}{\big[(r+r_0)+\im\, (a\cos\theta+l+\omega x_0)\big]^2}\,,
\end{align}
see also \eqref{phi_1_for_alpha=0} with \eqref{GP_trans}. In such a case, the two eigendirections of the electromagnetic field are \textit{both aligned} with the two (double degenerate) PNDs of the Weyl tensor.

Let us now employ the expressions (\ref{n_rel}), (\ref{eps_rel}), (\ref{b0tild}) to find the PD coefficients $n$, $\epsilon$, $k$. Taking the limit ${\alpha\rightarrow 0},~c=\mathrm{const.}$, we derive that the only possible finite solution is
\begin{align}
    n=l\,, \qquad
    \epsilon=1\,,\qquad
    \omega^2 k=a^2-l^2\,, \label{n-epsilon-k?alpha=0}
\end{align}
see also the expressions \eqref{n_GP}, \eqref{eps_GP}, \eqref{k_GP} for ${\alpha=0},~|c|=\mathrm{const}$.
Moreover, ${\tilde{\mathcal{P}}=1}$ because ${\tilde{a}_3=0=\tilde{a}_4}$ due to \eqref{tilda_ai} and \eqref{a_i-for-alpha=0}.

Finally, performing the shift \eqref{GP_trans_2} and introducing new mass and NUT parameters as
\begin{align}
    \tilde{m}:=m+r_0\,,\qquad \tilde{l}:=l+\omega\,x_0\,,
    \label{tilde-m}
\end{align}
we obtain the metric in the form \eqref{GP_metr_2} with ${\Omega=1=\tilde{\mathcal{P}}}$, that is

\newpage
\begin{align}
    \dd s^2 = &-\dfrac{\mathcal{Q}}{\varrho^2}
    \Big(\dd t-\big[a\sin^2\theta
    +2\tilde{l}\,(1-\cos\theta)\big]\dd\varphi\Big)^2
    +\dfrac{\varrho^2}{\mathcal{Q}}\,\dd \tilde{r}^2 \nonumber\\
    &+\varrho^2\dd\theta^2+\dfrac{\sin^2\theta}{\varrho^2}
    \Big(a\,\dd t-\big[\tilde{r}^2+(a+\tilde{l})^2\big]\dd\varphi\Big)^2,\label{GP_metric-alpha=0}
\end{align}
where
\begin{align}
    \varrho^2&=\tilde{r}^2+(a\cos\theta+\tilde{l})^2,\nonumber\\
    \mathcal{Q}&=a^2-\tilde{l}^2+e^2+g^2-2\tilde{m}\,\tilde{r}+\tilde{r}^2, \label{GP_metric-functions-alpha=0}\\
    \Phi_0 &= 0 = \Phi_2 \,,\qquad
    \Phi_1 = \dfrac{\tfrac{1}{2}(e+\im\, g)}{\big[\tilde{r}+\im\, (a\cos\theta+\tilde{l})\big]^2}\,. \nonumber
\end{align}

Actually, for the derivation of $\mathcal{Q}$ we have employed the identity
\begin{align}
\big(r_0^2+\omega^2 x_0^2\big) + 2 (m\,r_0 + l \omega\,x_0) = e^2 + g^2.
\end{align}
This can be proven by the direct evaluation of \eqref{x0r0-in-ABR}, in which the coefficients \eqref{defABnew} for ${\alpha=0},~|c|=\mathrm{const.}$ are simplified to
\begin{align}
A = \dfrac{4|c|^2}{1 - 4|c|^2}\,l\,,\qquad
B = \dfrac{4|c|^2}{1 - 4|c|^2}\,m\,,\qquad
R = \dfrac{4|c|^2}{1 - 4|c|^2}\,\sqrt{m^2+l^2}\,.\label{ABR_alph_0_lim}
\end{align}
Indeed,
\begin{align}
\big(r_0^2+\omega^2 x_0^2\big) + 2 (m\,r_0 + l \omega\,x_0)  &= \dfrac{8|c|^2}{(1 - 4|c|^2)^2}\,
  \big[ m^2+l^2 + \sqrt{m^2+l^2}\,(m \cos\beta + l \sin\beta) \big]\,,
\end{align}
which is the \emph{same expression} as for the identity \eqref{e2+g2=r2+omega2}, that is
\begin{align} \label{eg-identity}
    e^2+g^2=\dfrac{1}{4|c|^2}\big(r_0^2+\omega^2 x_0^2\big)\,.
\end{align}

The metric \eqref{GP_metric-alpha=0}, \eqref{GP_metric-functions-alpha=0} is exactly the \textit{usual form of the Kerr-Newman-NUT black hole} with an aligned electromagnetic field. This also demonstrates that turning off the acceleration parameter $\alpha$ while keeping $|c|=\mathrm{const.}$ also turns off the non-aligned part of the electromagnetic field $\Phi_0, \Phi_2$. The aligned part $\Phi_1$ is characterized by two parameters $e$ and~$g$, which are the electric and magnetic charges of the black hole. It justifies the introduction of these parameters in Section~\ref{sec_em_field}.

Let us discuss the role of the complex parameter $c$ in this subcase. As we mentioned, the non-aligned part vanishes in this limit. However, the parameter $|c|$ enters the expressions for $A$, $B$ and $R$ in (\ref{ABR_alph_0_lim}), and thus the charges $e$ and $g$. This allows to say that in this limit the parameter $c$ represents the charges of a black hole. However, generally this parameter also enters the non-aligned part of the electromagnetic field, so to properly interpret the role of $c$ in other special cases, a more careful analysis must be performed.

\subsection{No twist (${\omega=0}$): Van den Bergh-Carminati}
\label{section-no twist}

The \emph{static subcase} is not so straightforward, as one has to take the $\omega\rightarrow 0$  limit in the original metric (\ref{ds_alpha_omega}) while keeping $\omega n=\mathrm{const}$. Because of this complication, we postpone it for the separate work \cite{OP-prepar} in which we will show that there is a one-to-one correspondence between the static limit of our solution and the Van den Bergh-Carminati solution found in \cite{VandenBergh2020}, as given here by the metric (\ref{VandenBergh-metric}) that is equivalent to (\ref{orig_metr}).  Moreover, our case with non-zero $r_0$ corresponds to the ${hj\neq 0}$ subcase, while the case ${r_0=0}$ corresponds to ${h=0=j}$ subcase of \cite{VandenBergh2020}. In addition, note that this solution is not contained in the Alekseev-Garcia solution \cite{Alexeev1996} as it is of type I \cite{Ortaggio2018}.

\subsection{Special non-aligned electromagnetic field (${e=0=g}$)}
\label{section-e=0=g}

Let us recall (cf. Section~\ref{sec_em_field}) that the electromagnetic field is determined by 3 parameters, namely $|c|$ and two angular parameters $\beta, \gamma$ representing specific duality rotations \eqref{eg_another-form}, \eqref{r0-x0_another-form} which ``mix'' the charges $e,g$ and the related parameters $r_0, \omega\,x_0$. For the special choice ${\beta=\beta_0}$ given by \eqref{beta_cond-new}, that is for
\begin{align}
    \tan \beta_0=\dfrac{n\,-\alpha\omega\,m\,C}{m+\alpha\omega\,n\,C}\,, \label{beta_cond-again}
\end{align}
the coefficients $r_0$ and $\omega\,x_0$ are both zero, and this is equivalent to vanishing of both $e$ and~$g$,
\begin{align}
    r_0=0=\omega\,x_0 \qquad\Leftrightarrow\qquad e=0=g\,,
    \label{r0=x_0=0-e=0=g}
\end{align}
see \eqref{eg_def}, \eqref{r0x0-given-by-eg} for ${|c|\not=0}$.
This considerably simplifies the metric, preserving the existence of \emph{both aligned and non-aligned components} of the electromagnetic field ${\Phi_1\ne0}$, ${\Phi_0=\Phi_2\ne0}$.
Indeed, the metric \eqref{GP_metr} reduces to
\begin{align}
    \dd s^2=\dfrac{1}{\Omega^2}\Bigg[&-\dfrac{\mathcal{Q}}{\varrho^2}
    \Big(\dd t-\big[a\sin^2\theta + 2l\,(1-\cos\theta)\big]\dd\varphi\Big)^2
     +\dfrac{\varrho^2}{\mathcal{Q}}\,\dd r^2 \nonumber\\
    &+\dfrac{\varrho^2}{\tilde{\mathcal{P}}}\dd\theta^2+\dfrac{\tilde{\mathcal{P}}}{\varrho^2}\sin^2\theta
    \Big(a\,\dd t-\big[r^2+(a+l)^2\big]\dd\varphi\Big)^2\Bigg],\label{GP_metr_2-again}
\end{align}
where $\Omega^2$ is given by \eqref{Om2_rescl_GP},
\begin{align}
    \tilde{\mathcal{P}} & =1-\Big(\dfrac{a}{\omega}\,a_3+\dfrac{4al}{\omega^2}\,a_4\Big)\cos\theta-\dfrac{a^2}{\omega^2}\,a_4 \cos^2\theta\,, \nonumber\\
    \mathcal{Q}         & = b_0+b_1\,r+b_2\,r^2+b_3\,r^3+b_4\,r^4\,,\label{math-Qe=0=g}\\
    \varrho^2           & = r^2+(a\cos \theta+l)^2\,,  \nonumber
\end{align}
with the coefficients $a_i, b_i$ determined by \eqref{prim_param_6}, \eqref{prim_param_1}, and $n, \epsilon, k, C$ by \eqref{n_rel}--\eqref{a0=1}. Actually, this is the metric \eqref{GP_metr_2} because \eqref{GP_trans_2} simplifies to trivial relations
\begin{equation}
 r=\tilde{r}\,, \qquad  l=\tilde{l} \,.
 \label{r,l}
\end{equation}

In fact, this metric form is the same as for the Griffiths-Podolsk\'y form of type D black holes, see Eq.~(16.18) in \cite{GriffithsPodolsky:2009}. But the Maxwell field is now extended to non-aligned components \eqref{phi_0-GP}, \eqref{phi_1-GP},
\begin{align}
\Phi_0&=\Phi_2=\,\dfrac{\alpha c}{\Omega}\,\dfrac{a}{\omega}\,
        \dfrac{\sqrt{\tilde{\mathcal{P}}\mathcal{Q}}\,\sin\theta}
        {r+\im\,(a\cos\theta+l)}\,,\label{phi_0-GP-eg=0} \\
\Phi_1&=\dfrac{1}{4\alpha\,\bar{c}\,\sin\theta}\,\dfrac{\omega}{a}\,
    \dfrac{\Omega^2}{[r+\im\,(a\cos\theta+l)]^2} \Big[
     (a\cos\theta+l)^2\Big(\dfrac{-\Omega_{,r}}{a\cos\theta+l}\Big)_{,\theta}
     +\im\, r^2\Big(\dfrac{\Omega_{,\theta}}{r}\Big)_{,r} \,\Big].\label{phi_1-GPeg=0}
\end{align}
Such electromagnetic field is described by \eqref{A-6} which due to the relation \eqref{r,l} reads
\begin{align}
    \mathbf{A} = \dfrac{1}{4\alpha\,\bar{c}}\,\dfrac{\omega}{a}\Big[\,
    &\Omega_{,r}\,\dfrac{ a\,\dd t-\big[\,{r}^{\,2}+(a+ l)^2\big]\dd\varphi }
        {r+\im\,(a\cos\theta+ l)}
            \label{A-7}\\
   +&\dfrac{\im\,\Omega_{,\theta}}{\sin\theta}\,
     \dfrac{ \dd t - \big[a\sin^2\theta+2 l (1-\cos\theta)\big]\dd\varphi}
       {r+\im\,(a\cos\theta+ l)} + \Omega\,\dd\varphi\, \Big]  + \mathbf{A}_0\,.\nonumber
\end{align}

This $e=0=g$ subcase is important for the physical interpretation of the new spacetime, as now it is easier to set some of the remaining parameters to zero, and to understand their role (which is rather complicated if $e,g$ are nonzero). This is done in the following Subsections.

\subsection{No acceleration ${(\alpha=0},~|c|=\mathrm{const.},~e=0=g)$: Kerr-NUT}

We can employ the general result derived in Section~\ref{gen_no_accel}, namely that in the ${\alpha\rightarrow 0}$, ${|c|=\mathrm{const.}}$ limit the electromagnetic field becomes aligned. Moreover, it follows from \eqref{Phi_2-for-alpha=0} that for ${e=0=g}$ the field \emph{completely vanishes}, ${\Phi_0 = \Phi_2 =0}$, ${\Phi_1 = 0}$. The corresponding metric \eqref{GP_metric-alpha=0} is
\begin{align}
    \dd s^2 = &-\dfrac{\mathcal{Q}}{\varrho^2}
    \Big(\dd t-\big[a\sin^2\theta
    +2l\,(1-\cos\theta)\big]\dd\varphi\Big)^2
    +\dfrac{\varrho^2}{\mathcal{Q}}\,\dd r^2 \nonumber\\
    &+\varrho^2\dd\theta^2+\dfrac{\sin^2\theta}{\varrho^2}
    \Big(a\,\dd t-\big[r^2+(a+l)^2\big]\dd\varphi\Big)^2,\label{GP_metric-alpha=0,eg=0}
\end{align}
where
\begin{align}
    \varrho^2 =r^2+(a\cos\theta+l)^2,\qquad
    \mathcal{Q}=a^2-l^2-2m\,r+r^2, \label{GP_metric-functions-alpha=0,eg=0}
\end{align}
which is the usual form of the Kerr-NUT black hole, see Eq.~(16.23) in \cite{GriffithsPodolsky:2009}.

\subsection{No NUT ${(l=0,~e=0=g)}$: New uncharged black holes}
\label{subsection e=0=g and l=0}

In this case, the metric \eqref{GP_metr_2-again} becomes
\begin{align}
    \dd s^2=\dfrac{1}{\Omega^2}\Bigg[&-\dfrac{\mathcal{Q}}{\varrho^2}
    \Big(\dd t-a\sin^2\theta\,\dd\varphi\Big)^2
    +\dfrac{\varrho^2}{\mathcal{Q}}\,\dd r^2
    +\dfrac{\varrho^2}{\tilde{\mathcal{P}}}\dd\theta^2+\dfrac{\tilde{\mathcal{P}}}{\varrho^2}\sin^2\theta\,
    \Big(a\,\dd t-(r^2+a^2)\,\dd\varphi\Big)^2\Bigg],\label{GP_e=0=g,l=0}
\end{align}
where
\begin{align}
  \Omega^2  &= \Big(1-\dfrac{\alpha}{\omega}\,a\,r\cos\theta\Big)^2 \nonumber\\
     & \hspace{4mm} +2\alpha^2C\Big[ \,r^2
        +\Big(2\alpha^2C-\frac{\epsilon}{\omega^2k}\Big)\,a^2r^2\cos^2\theta -a^2\cos^2\theta \,\Big] \label{Om2_e=0=g,l=0}\\
     & \hspace{4mm} + \dfrac{4\alpha^2 C}{\omega^2 k}
         \Big(\,\frac{m+\alpha \omega\, Cn}{1+\alpha^2\omega^2C^2}\,a^2 r\cos^2\theta
             +\frac{n-\alpha \omega\, Cm}{1+\alpha^2\omega^2C^2}\,a\,r^2\cos\theta\Big), \nonumber\\
\tilde{\mathcal{P}} & =1-\dfrac{a}{\omega}\,a_3\cos\theta-\dfrac{a^2}{\omega^2}\,a_4 \cos^2\theta\,,
         \label{tilde_P-e=0=g,l=0}\\
\mathcal{Q} & = b_0+b_1\,r+b_2\,r^2+b_3\,r^3+b_4\,r^4\,,\label{math-Q-e=0=g,l=0}\\
\varrho^2   & = r^2+a^2\cos^2\theta\,. \label{rho2_-e=0=g,l=0}
\end{align}

As argued in \cite{Ovcharenko2024}, it is possible to choose any value of the twist parameter $\omega$ by a suitable rescaling of the acceleration parameter $\alpha$. In this particular case, the most natural choice is
\begin{align}
\omega=a\,, \label{omega=a}
\end{align}
because this simplifies the metric functions. Indeed, evaluating the complicated square roots \eqref{s1-and-s2} for ${l=0}$ we get very simple expressions ${I_1=1-\alpha^2a^2C}$, ${I_2=1-2\alpha^2a^2C}$, so that \eqref{b0tild} with the upper (minus) sign gives ${\tilde{a}_0=k}$. Equations \eqref{a0=1} and  \eqref{C def-again} thus imply
\begin{align}
    k =1\,,\qquad
    C = 2 |c|^2\,. \label{kC}
\end{align}
To express (\ref{n_rel}) and (\ref{eps_rel}), we must expand $I_1$ and $I_2$ to the \emph{second order} in $l$ and then perform the limit ${l\to0}$. Nevertheless, even in this more involved case we obtain modest explicit relations
\begin{align}
    n&=-\alpha a m\,\dfrac{1}{I_1} \big(1-2|c|^2 I_2\big) \,, \nonumber\\
\epsilon&=1-4\alpha^2m^2|c|^2 \dfrac{I_2}{I_1^2}-\alpha^2a^2 \dfrac{1}{I_2}\big(1-4|c|^2 I_2\big),
\end{align}
where
\begin{align}
    I_1 &= 1-2\alpha^2 a^2|c|^2\,,\nonumber\\[2mm]
    I_2 &= 1-4\alpha^2 a^2|c|^2\,,\label{defI1I2}
\end{align}
so that
\begin{align}
 \frac{m+\alpha \omega\, Cn}{1+\alpha^2\omega^2C^2} = m\,\frac{I_2}{I_1}\,,\qquad\qquad
 \frac{n-\alpha \omega\, Cm}{1+\alpha^2\omega^2C^2} = -\alpha\,a\,m\,\frac{1}{I_1}\,.
\end{align}
Using these very simple expressions, the coefficients \eqref{prim_param_6}, \eqref{prim_param_1} reduce to
\begin{align}
    a_3&= 2\alpha m \,\dfrac{1}{I_1}\,, \nonumber\\
    a_4&= - \alpha^2a^2\dfrac{1}{I_2}\big(1-4|c|^2I_2\big)-4\alpha^2m^2|c|^2\frac{I_2}{I_1^2}\,, \label{prim_param_6-e=0=g,l=0}
\end{align}
(actually, ${a_4=\epsilon-1}$, ${b_2=\epsilon}$), and
\begin{align}
    b_0&= a^2\,, \nonumber\\
    b_1&=-2m\,\dfrac{I_2}{I_1}\,, \nonumber\\
    b_2&= \Big(1-4\alpha^2 m^2 |c|^2 \dfrac{I_2}{I_1^2}\Big)-\alpha^2a^2 \dfrac{1}{I_2}\big(1-4|c|^2 I_2\big),\label{prim_param_1-e=0=g,l=0}\\
    b_3&= 2\alpha^2m\,\dfrac{1}{I_1}\big(1-4|c|^2 I_2\big), \nonumber\\
    b_4&= -\alpha^2\frac{1}{I_2}\Big(1-4\alpha^2 m^2|c|^2 \dfrac{I_2}{I_1^2}\Big)\big(1-4|c|^2 I_2\big)\,.
    \nonumber
\end{align}

Therefore, the \emph{quadratic} metric function $\tilde{\mathcal{P}}(\cos\theta)$ in \eqref{GP_e=0=g,l=0} reads
\begin{align}
    \tilde{\mathcal{P}} = 1-2\alpha m \,\dfrac{1}{I_1}\cos\theta
        +\alpha^2\Big[a^2\,\dfrac{1-4|c|^2I_2}{I_2}
                 + 4m^2 |c|^2 \dfrac{I_2}{I_1^2} \Big]\cos^2\theta\,,
\end{align}
and (miraculously) the \emph{quartic} $\mathcal{Q}(r)$ \emph{is factorized} as
\begin{align}
\mathcal{Q}=\Big[\,a^2-2m\,\dfrac{I_2}{I_1}\,r+\Big(1-4\alpha^2m^2|c|^2\,\dfrac{I_2}{I_1^2}\Big)\,r^2\,\Big]
    \,\Big[\,1-\alpha^2\,\dfrac{1-4|c|^2 I_2}{I_2}\,r^2\,\Big]\,.
\end{align}

This factorization enables us to easily calculate the position of the \emph{four horizons} of this new family of black holes with non-aligned Maxwell field, namely
\begin{align}
r_b^{\pm}&= \dfrac{m\,I_2 \pm \sqrt{m^2I_2^2-a^2(I_1^2-4\alpha^2m^2|c|^2 I_2)}}
    {I_1^2-4\alpha^2m^2|c|^2I_2} \,I_1 \,,  \label{rb}\\[1mm]
r_a^{\pm}&=\pm \dfrac{1}{\alpha}\,\sqrt{\dfrac{I_2}{1-4|c|^2 I_2}}\,. \label{ra}
\end{align}
The former are \emph{two black hole horizons}, whereas the later are \emph{two acceleration horizons}.
\vspace{2mm}

The conformal factor in (\ref{GP_e=0=g,l=0}) is now given by a compact expression
\begin{align} \label{Omega-e=0=g,l=0}
    \Omega^2=\,\big(1-\alpha r\cos\theta\big)^2
   + 4\alpha^2|c|^2 \Big( J\,r^2 + 2m\,\dfrac{I_2}{I_1}\,r\cos^2\theta - a^2\cos^2\theta \Big)\,,
\end{align}
where
\begin{align} \label{Omega-e=0=g,l=0XXX}
    J(\cos\theta)& := \,\tilde{\mathcal{P}}-(1-4\alpha^2a^2|c|^2)\cos^2\theta \nonumber\\
    &\,\, = \,\sin^2\theta-2\alpha m \,\dfrac{1}{I_1}\cos\theta
    +\alpha^2\Big(a^2\dfrac{1}{I_2} + 4m^2|c|^2 \dfrac{I_2}{I_1^2} \Big)\cos^2\theta\,.
\end{align}
It reduces to ${\,\Omega=1-\alpha r\cos\theta\,}$ when ${c=0}$.

The only Weyl curvature NP component is given by the scalar
\begin{align}\label{psi2-for l=0}
    \Psi_2 = - m\,\dfrac{\Omega^2}{I_1^2}\bigg[&
     \dfrac{I_1}{(r+\im\,a \cos\theta)^3}\Big(I_2-\alpha (r\cos\theta+\im\,a)
        +\alpha^2 (1-4|c|^2I_2)\,\im \,a\,r \cos\theta\Big)\nonumber\\
    &+\dfrac{4\alpha^2 m|c|^2r^2\cos^2\theta}{(r^2+a^2\cos^2\theta)(r+\im\, a \cos\theta)^2}\bigg].
\end{align}
The \emph{curvature singularity} occurs at ${r=0}$, but \emph{only if also} ${\theta=\frac{\pi}{2}}$ and $a\neq 0$. It thus has a \emph{ring structure}, similarly as in the Kerr spacetime. When ${\theta\ne\frac{\pi}{2}}$, it is possible to reach the region ${r<0}$.

The potential of the Maxwell field (\ref{A-7}) is simplified to
\begin{align}
    \mathbf{A}=\dfrac{1}{4\alpha\,\bar{c}}\,\Big[\,
    &\Omega_{,r}\,\dfrac{ a\,\dd t-({r}^{\,2}+a^2)\dd\varphi }
        {r+\im\,a\cos\theta}
        +\dfrac{\im\,\Omega_{,\theta}}{\sin\theta}\,
     \dfrac{ \dd t - a\sin^2\theta\,\dd\varphi}
       {r+\im\,a\cos\theta} + \Omega\,\dd\varphi\, \Big]  + \mathbf{A}_0\,.\nonumber
\end{align}

The \emph{nonaligned part} of the electromagnetic field is given by
\begin{align}
    \Phi_0=\Phi_2 = \alpha \,c\,\,\dfrac{1}{\Omega}\,
    \dfrac{\sqrt{\tilde{\mathcal{P}}\mathcal{Q}}\,\sin\theta }{r+\im\,a\cos\theta}\,.
        \label{Phi0,1=0,e=0=g}
\end{align}
Interestingly, it \emph{vanishes on the horizons} (where ${\mathcal{Q}=0}$) and also \emph{along the axis of symmetry} (at ${\theta=0, \pi}$).
The aligned component is more complicated, namely
\begin{align}
    \Phi_1 = \alpha\,c \,\,\,
    \dfrac{B_0+B_1\,r+B_2\,r^2+B_3\,r^3}{I_1^3I_2\,\Omega\,(r+\im\,a\cos\theta)^2}\,,
     \label{Phi1,l=0,e=0=g}
\end{align}
in which $B_i(\cos\theta)$ are the following functions, independent of $r$,
\begin{align}
B_0 &= a\, I_1^2I_2\cos\theta\,\big[ mI_2\,\cos\theta
   -\alpha a^2I_1(1-\im\,4\alpha a|c|^2)\cos^2\theta - \im\,a\,I_1 \big],\nonumber\\
B_1 &= a\, I_1\,\big[\!-I_1^2I_2(1+\cos^2\theta) + \alpha m\,I_2^2\cos^2\theta \,(4\alpha m |c|^2+I_1\cos\theta) \nonumber\\
 &\hspace{14mm}  + \alpha^2a^2 I_1\cos^2\theta\,(I_1+8|c|^2I_1I_2-8\alpha m |c|^2I_2\cos\theta)\,\nonumber\\
 &\hspace{14mm}  + \im\,3\alpha a\, I_1I_2\cos^2\theta\,(I_1-4\alpha m |c|^2I_2\cos\theta)\big],\label{Bi}\\
B_2 &= I_1\,\big[3\alpha a\,I_2 (I_1-\alpha m\cos\theta)(I_1-4\alpha m |c|^2I_2\cos\theta)\cos\theta-\im\,4\alpha^4a^4|c|^2 I_1^2\cos^3\theta - \im\,\alpha^2a^2 D \cos\theta\nonumber\\
&\hspace{14mm}  + \im\,I_2\,\big(I_1^2\cos\theta-4\alpha^2m^2|c|^2I_2\cos\theta\,(1-2I_2\cos^2\theta)+ \alpha m I_1(1-3I_2\cos^2\theta)\big)\nonumber \big],
 \nonumber\\
B_3 &= -\alpha\,\big[\alpha a I_2 (I_1-\alpha m \cos\theta)\big((I_1^2+16\alpha^2m^2|c|^4I_2)\cos^2\theta
                 - 4|c|^2 I_1(2\alpha m \cos\theta-I_1\sin^2\theta)\big)\nonumber\\
 &\hspace{0mm}  +4\alpha^3a^3|c|^2I_1^2 (I_1-\alpha m\cos\theta)\cos^2\theta
-\im\,4\alpha^3a^2m|c|^2 I_1^2I_2\cos^3\theta +\im\,I_2\big(I_1^3- \alpha m\,D \cos\theta \big)\big] ,\nonumber
\end{align}
where
\begin{align}
    D=I_1^2\,\big[1+8|c|^2I_2+ (1-4|c|^2)I_2 \cos^2\theta\big]+4\alpha m |c|^2 I_2 \cos\theta (4\alpha m |c|^2 I_2 \cos\theta-3 I_1).\end{align}

Now, after we described the general setting for this interesting class of accelerating uncharged black holes, let us consider several possible limits.

\subsubsection{Kerr}

In the ${\alpha=0},~{|c|=\mathrm{const.}}$ subcase we get ${\Omega=1}$, ${\tilde{\mathcal{P}}=1}$,
${\mathcal{Q}=a^2-2m\,r+\,r^2}$ and ${\Phi_0 = \Phi_2 =0}$, ${\Phi_1 = 0}$. The corresponding metric is simply the \emph{Kerr black hole} spacetime with two horizons located at
\begin{align}
 r_b^{\pm}=m\pm\sqrt{m^2-a^2} \,,
\end{align}
which has the standard form \eqref{GP_metric-alpha=0,eg=0} for ${l=0}$.

\subsubsection{Kerr with acceleration}

In the ${c=0},~{\alpha=\mathrm{const.}}$ subcase the coefficients (\ref{defI1I2}) become ${I_1=1=I_2}$. The metric functions in  \eqref{GP_e=0=g,l=0} thus reduce to the form ${\,\Omega=1-\alpha r\cos\theta\,}$, ${\tilde{\mathcal{P}} = 1-2\alpha m \,\cos\theta + \alpha^2a^2\cos^2\theta}$, ${\mathcal{Q}=(a^2-2m\,r+r^2)(1-\alpha^2\,r^2)}$, and expressions \eqref{rb}, \eqref{ra} read
\begin{align}
 r_b^{\pm}=m\pm\sqrt{m^2-a^2} \,,\qquad
 r_a^{\pm}=\pm \dfrac{1}{\alpha}   \,.
\end{align}
These are the positions of the four horizons for usual \emph{accelerating Kerr black} hole (rotating C-metric solution) without electromagnetic field, see \cite{Podolsk2021}.

\subsubsection{Kerr in a magnetic field}

In the special limit ${\alpha\rightarrow 0}$, ${|c|\rightarrow \infty}$, such that ${2\alpha|c|\equiv B=\mathrm{const.}}$, the coefficients $I_1$ and~$I_2$ become
\begin{align}
    I_1=1-\frac{1}{2}B^2a^2\,,\qquad I_2=1-B^2a^2\,.
\end{align}
The metric functions in (\ref{GP_e=0=g,l=0}) simplify to
\begin{align}
    \tilde{\mathcal{P}}&=1+B^2\Big(m^2\dfrac{I_2}{I_1^2}-a^2\Big)\cos^2\theta\,,\label{P_Kerr-BR}\\
    \mathcal{Q}&=(1+B^2r^2)\Delta\,,\\[2mm]
    \Omega^2&=(1+B^2r^2)-B^2\Delta\cos^2\theta\,,\label{Om2_Kerr-BR}
\end{align}
where
\begin{align}
    \Delta=\Big(1-B^2 m^2\dfrac{I_2}{I_1^2}\Big)r^2-2m \dfrac{I_2}{I_1}+a^2\,.
\end{align}
The two horizons (\ref{rb}) are located at
\begin{align}
    r_b^{\pm}= \dfrac{m \,I_2\pm \sqrt{m^2 I_2-a^2I_1^2}}{I_1^2 -B^2m^2I_2}\,I_1\,.
\end{align}

This solution is precisely the new spacetime presented in \cite{Kerr-BR}, which represents \textit{the Kerr black hole immersed in an external uniform magnetic field}. Such physical interpretation comes from the fact that for ${B=0}$ we recover the standard Kerr metric. In the ${m=0}$ case, after a proper coordinate transformation, one obtains the Bertotti-Robinson spacetime (see Section II.B in \cite{Kerr-BR}), which is known to describe a uniform electromagnetic field \cite{GriffithsPodolsky:2009}. The parameter $B$ in this solution stands for the value of the magnetic field. The parameter~$\gamma$, entering expressions for $\Phi_0,~\Phi_1,~\Phi_2$ (see Eqs. (2.11) and (2.12) in \cite{Kerr-BR}) represents the duality rotation between the electric and magnetic field. Interestingly, the magnetic field is weakened and expelled away in the equatorial plane, exhibiting the Meissner effect.

\section{Summary of the particular cases and structure of the new class}

We have presented a new solution, and analyzed its various main subcases. Now let us summarize the role of each of its physical parameters.

\begin{itemize}
    \item The parameters $m,~a,~l,~\alpha$.
\end{itemize}
Meaning of these parameters seems to be more or less straightforward in various subcases: $m$ is related to the mass of a black hole, $a$ to its Kerr-like parameter, $l$ to the NUT twist parameter, and $\alpha$ to the acceleration. However, one has to be careful with the precise interpretation in the general case. This warning is necessary because even for the classic Pleba\'nski-Demia\'nski spacetime, in the most general case there exist various metric forms (see \cite{Ovcharenko2024,Ovcharenko2025}), and relation between their physical parameters is complicated. This may lead to some incorrect statements while considering various special cases. We expect the same situation to appear here.
\begin{itemize}
    \item The parameter $|c|$
\end{itemize}
This is the new parameter in our class of solutions. Its physical interpretation is not straightforward. However, we can say that it is related \textit{both} to the (electric and magnetic) charges of a black hole, and to the strength of the external electromagnetic field. This interpretation comes from two limits. If $\alpha=0,~|c|=\mathrm{const.}$, then (as we showed in Section \ref{gen_no_accel}) the non-aligned part disappears and the strength of the aligned part is related to the charges $e$ and $g$ that depend on $|c|$. Thus, $|c|$ is related to the charges of a black hole. On the other hand, if $\alpha\rightarrow 0$ and $2\alpha|c|\equiv B=\mathrm{const.}$ with $\beta=\beta_c$ (see Section \ref{subsection e=0=g and l=0}), then the corresponding solution is the Kerr black hole immersed in an external magnetic (or electric) field, and in this case $|c|$ is related to the value $B$ of this external electromagnetic field. Moreover, if $|c|=0$ while all other parameters are kept finite, the electromagnetic field vanishes. This means that $|c|$ is \textit{solely related to the electromagnetic field}.

\begin{itemize}
    \item The parameters $e$ and $g$
\end{itemize}
These are \textit{auxiliary} parameters, introduced in (\ref{eg_def}), and they cannot be considered as independent ones (instead of them it is possible, without loss of generality, to use only the parameters $|c|$ and $\beta$). Their interpretation is based on the fact that in the limit ${\alpha=0}$, ${=\mathrm{const.}}$ they are equal to the electric and magnetic charges of the Kerr-Newman-NUT black hole itself. Thus $e$ and $g$ can be thought of as \textit{related to the physical charges of a black hole}. In the general case, however, one should calculate the corresponding fluxes of electric and magnetic fields to properly identify them, but this lies beyond the scope of this work.

\begin{itemize}
    \item The parameter $\beta$
\end{itemize}
This parameter appears in the relations (\ref{x0r0-in-ABR}), and represents a duality rotation between the electric and magnetic charges of a black hole itself, as $r_0$ and $x_0$ are related to the charges $e$ and $g$ by (\ref{r0x0-given-by-eg}).

\begin{itemize}
    \item The parameter $\gamma$
\end{itemize}
The role of this parameter seems to be quite straightforward because it is the phase of the complex parameter $c$, see (\ref{def_gamma}). It thus directly enters in the expressions (\ref{phi_0-GP}), (\ref{phi_1-GP}) for $\Phi_0$, $\Phi_1$, $\Phi_2$, and \textit{can be interpreted} as the \textit{duality rotation parameter} of the external electromagnetic field.

\vspace{2mm}

In conclusion, it is important to emphasize that the above parametrization, in which we have presented this large class of new solutions, need not be the best one for the physical interpretation, and additional work in finding more convenient parametrization can/must be done. Nevertheless, from the analysis of the special cases we may conclude that the most general solution can be interpreted as representing a massive, charged, accelerating black holes with the Kerr and NUT twist parameters, immersed into an external electromagnetic field. This external field is the reason why the electromagnetic field is not aligned, as a distinctive feature from the Kerr-Newman family, generalized in the whole Pleba\'nski-Demia\'nski class. Interestingly enough, the gravitational field remains of algebraic type D.

\newpage

\section{Main physical properties of the general class}
\label{Sec-Physics}

The metric functions in the most general case (\ref{GP_metr})--(\ref{Om2_rescl_GP}) are quite complicated. Explicit investigation has been done only in particular cases (as discussed above). However, the most general case also requires physical analysis, which we are going to do now using also numerical methods. Moreover, the new solution differs from the Pleba\'nski-Demia\'nski class in the form of the metric functions $\mathcal{Q}$, $\tilde{\mathcal{P}}$, $\Omega$, and in the non-trivial structure of the electromagnetic field. In this section, we investigate both of these features.

\subsection{Positions of the horizons}

Here we aim to investigate how the horizons of this new class of black holes depend on the key parameter $|c|$. For this, we employ the form of the metric (\ref{GP_metr}) with $\Omega^2(r,\theta)$ given by (\ref{Om2_rescl_GP}), the function $\mathcal{Q}(r)$ is given by (\ref{math-Q}), the function $\tilde{\mathcal{P}}$ is given by (\ref{tilde_P}) with $\tilde{a}_3$ and $\tilde{a}_4$ determined by (\ref{tilda_ai}), the parameters $r_0$ and $x_0$ are given by \eqref{x0r0-in-ABR}, and $n,~\epsilon$, and $k$, entering \eqref{prim_param_6}--\eqref{prim_param_1}, are related to the physical parameters by (\ref{n_rel}), (\ref{eps_rel}), and (\ref{b0tild}), respectively. We choose this form of the metric because the metric function $\mathcal{Q}$ does not depend on the parameter $\beta$, so our results made in this Subsection will be valid for all $\beta$.

First of all, we have to solve Eqs.~(\ref{n_rel}), (\ref{eps_rel}), (\ref{b0tild}) for $k,~n,~\epsilon$. Their dependence on~$|c|$ is presented on the plots in Fig.~\ref{kne_plot} in Section \ref{Section_GP_form}. Generally, there are up to 5 roots. However, only one of them gives a finite real limit of $k,~n,~\epsilon$ as ${c\rightarrow 0}$ (namely the solid lines denoted as I in Fig. \ref{kne_plot}).

\begin{figure}
    \centering
\includegraphics[width=0.85\linewidth]{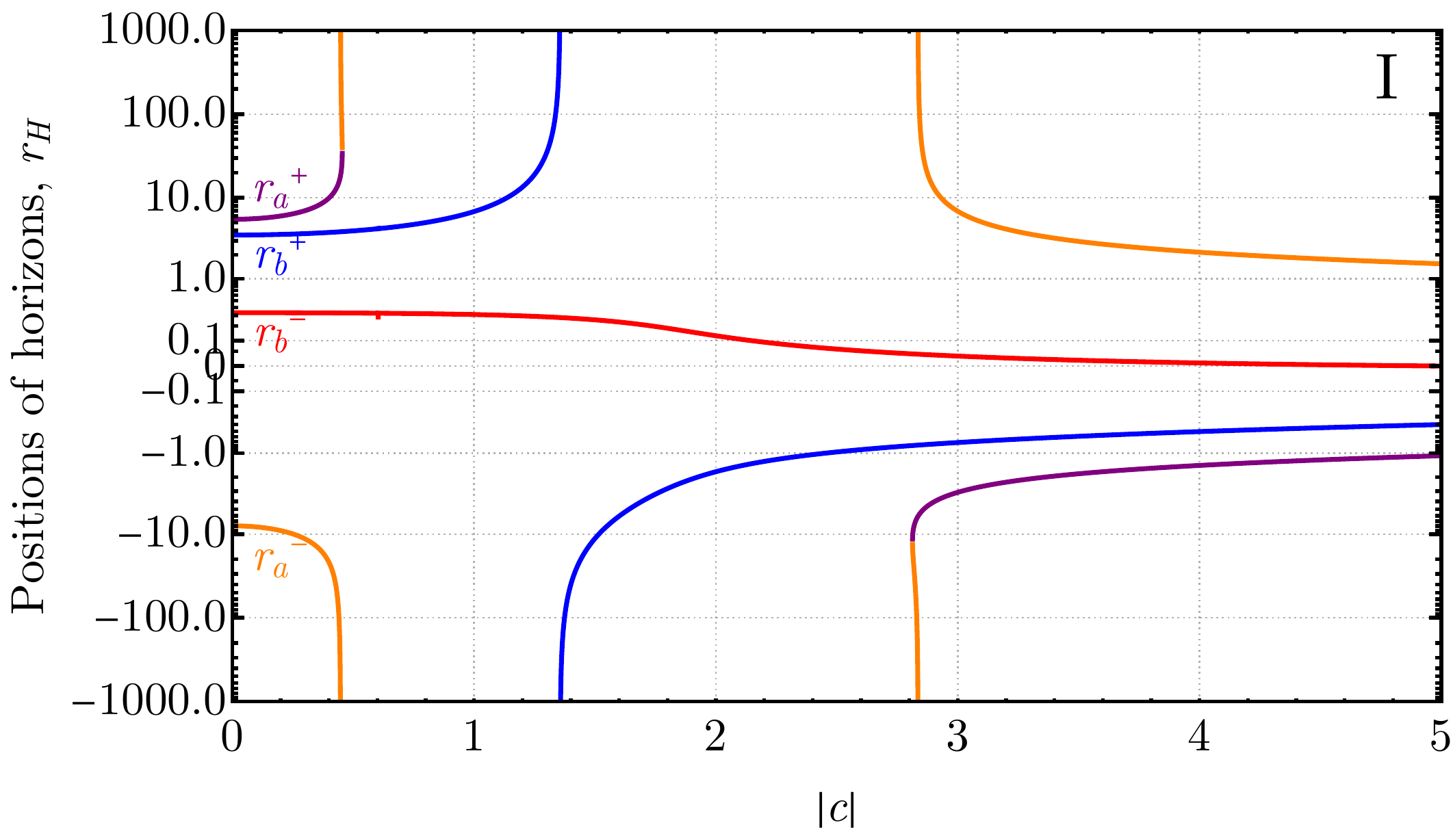}
    \caption{Plots showing the positions of horizons depending on $|c|$ for the branch~I of Fig~\ref{kne_plot} (plotted in symmetric logarithmic scale). Different colors show different horizons, namely blue and red curves represent the outer and inner black hole horizons $r_b^{\pm}$, respectively, while the purple and orange curves represent the outer and inner acceleration horizons $r_a^{\pm}$ (notations are taken from \cite{Podolsk2021,Podolsk2023}). The parameters employed here are ${m=2.2},~{a=1.1},~{l=0.2},~{\alpha=0.14},~{\omega=1}$.}
    \label{roots_plot}
\end{figure}

This root (the curves~I in Fig.~\ref{kne_plot}) in the ${c\rightarrow 0}$ limit corresponds to the Pleba\'nski-Demia\'nski solution, with the unique form \eqref{k_GP}--\eqref{eps_GP}. The corresponding \textit{positions of the horizons}, given by the roots $\mathcal{Q}(r)=0$ determined by \eqref{math-Q} and \eqref{prim_param_1}, are shown in Fig.~\ref{roots_plot}. First of all, we note that, as in the case of the Pleba\'nski-Demia\'nski solution, this more general solution also has 4 roots. However, in the PD case it is known that two of the roots do not depend on charge (see \cite{Podolsk2021}), namely the acceleration horizons $r_a^{\pm}$. In our case it appears that they depend on the parameter $|c|$ (these are the orange and purple curves in Fig.~\ref{roots_plot}). The second difference is that in the Pleba\'nski-Demia\'nski spacetime there are two black hole horizons at ${r_b^{\pm}=m\pm \sqrt{m^2+l^2-a^2-e^2-g^2}}$, see \cite{Podolsk2021}. As ${e^2+g^2}$ increases, these horizons converge and merge to one extremal horizon at ${e^2+g^2=m^2+l^2-a^2}$ located at $r_e=m$. In our case of the non-aligned electromagnetic field, these two roots are also present (the red and blue curves). However, instead of converging with an increase of the parameter $|c|$, they \textit{diverge}.

\begin{figure}
    \centering
\includegraphics[width=1\linewidth]{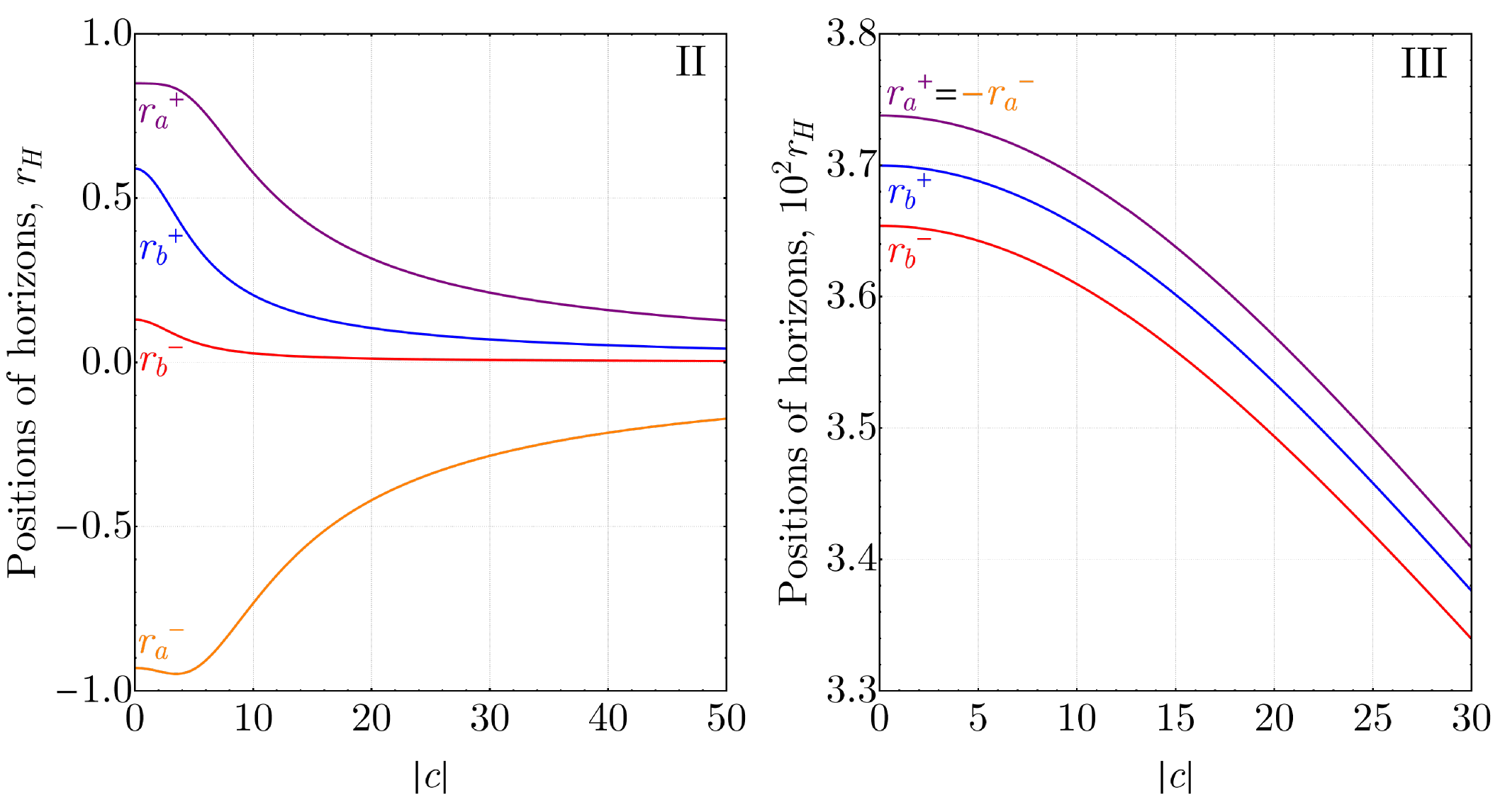}
    \caption{Plots showing the positions of horizons depending on $|c|$ for the branches~II~and~III of Fig.~\ref{kne_plot}. Different colors show different horizons, namely blue and red curves represent the outer and inner black hole horizons $r_b^{\pm}$, while the purple and orange curves represent the outer and inner acceleration horizons $r_a^{\pm}$. Notice a nice and uniform ordering $r_a^+>r_b^+>r_b^->r_a^-$ for all values of $|c|$. The parameters are $m=2.2,~a=1.1,~{l=0.2},~\alpha=0.14,~\omega=1$.}
    \label{roots_plot_2}
\end{figure}

Moreover, there are distinct 4 roots for $k,~n,~\epsilon$, as depicted in Fig.~\ref{kne_plot}. The branches IV and V give rise to the function $\mathcal{Q}$ that does not have any real roots for a given set of parameters, so we will not consider them here because these spacetimes \emph{have no horizons}. Solutions for the branches II and III are drawn in Fig.~\ref{roots_plot_2}. Note that both these branches give \textit{finite} positions of the 4 horizons in the limit $c\rightarrow 0$, despite the fact that the PD-like coefficients $k,~n,~\epsilon$ diverge in this limit (see Fig.~\ref{kne_plot}). However, as can be shown by the analysis of the electromagnetic field, these branches possess non-trivial poles. Such configurations are not physically expected, and thus the physical relevance of the branches II and III is questionable.

\vspace{5mm}

\subsection{Cosmic strings at ${\theta=0,\pi}$ causing the acceleration}
\label{subsec:strings}

The metric \eqref{GP_metr_2}, which is equivalent to \eqref{GP_metr}, is convenient for explicit analysis of the \emph{regularity of the poles/axes} located at ${\theta=0}$ and ${\theta=\pi}$, respectively. The spatial axes of symmetry are associated with the Killing vector field $\partial_\varphi$, and identified as zeros of the function $\sin\theta$ in the metric \eqref{GP_metr_2}. The range of the spatial coordinate~$\theta$ is thus constrained to ${\theta\in[0,\pi]}$.

Apart from 7 physical parameters in the new class of solutions, namely $m, a, l, \alpha, |c|, \beta, \gamma$, there is also \emph{eighth free parameter}, namely the \emph{conicity~$C$} hidden in the range of the angular coordinate
\begin{equation}
\varphi\in[0,2\pi C)\,,
 \label{conicity}
\end{equation}
which has not yet been specified. We will now demonstrate its physical meaning by relating it to \emph{deficit (or excess) angles} of the \emph{cosmic strings (or struts)}. Their internal tension is the \emph{physical source of the acceleration of the black holes}. These are basically topological defects associated with \emph{conical singularities} around the two distinct axes.

Let us start with investigation of the \emph{first axis of symmetry} ${\theta=0}$ in the metric (\ref{GP_metr_2}). Consider a small circle around it given by ${\theta=\hbox{const.}}$, with the range of $\varphi$ given by \eqref{conicity}, assuming fixed $t$ and~$r$. The invariant length of its \emph{circumference} is ${\int_0^{2\pi C}\!\! \sqrt{g_{\varphi\varphi}}\, \dd\varphi}$, while its \emph{radius} is ${\int_0^{\theta}\! \sqrt{g_{\theta\theta}}\, \dd\theta}$. The axis is regular if their fraction in the limit ${\theta\to 0}$  is equal to ${2\pi}$. However, in general we obtain
\begin{equation}
 f_0 := \lim_{\theta\to0} \frac{\hbox{circumference}}{\hbox{radius}}
 =\lim_{\theta\to0} \frac{2\pi C \sqrt{g_{\varphi\varphi}} }{ \theta\,\sqrt{g_{\theta\theta}}}  \,.
 \label{Accel-con0}
\end{equation}
For the metric \eqref{GP_metr_2}, the relevant metric functions are
\begin{align}
 g_{\varphi\varphi} &= \frac{1}{\Omega^2\varrho^2}\,
   \Big[\, \tilde{\mathcal{P}}\sin^2\!\theta\, \big[\tilde{r}^{\,2}+(a+\tilde{l}\,)^2\big]^2
   - \mathcal{Q} \big[a\sin^2\theta +2\tilde{l}(1-\cos\theta)\big]^2\,\Big],\nonumber\\[2mm]
 g_{\theta\theta} &= \frac{\varrho^2}{\Omega^2\, \tilde{\mathcal{P}}}\,,\hbox{\qquad where\quad}
    \varrho^2=\tilde{r}^2+(a\cos\theta+\tilde{l}\,)^2.
\end{align}
For small values of $\theta$, the second term in $g_{\varphi\varphi}$ proportional to $\mathcal{Q}$ becomes negligible compared to the first term proportional to $\tilde{\mathcal{P}}$, so that we obtain ${ g_{\varphi\varphi} \approx \tilde{\mathcal{P}} \big[\tilde{r}^{\,2}+(a+\tilde{l}\,)^2\big]^2\,\dfrac{\theta^2}{\Omega^2\varrho^2}}$.
Evaluation of the limit (\ref{Accel-con0}) using \eqref{tilde_P} gives
\begin{equation}
 f_0 =  2\pi C\,\tilde{\mathcal{P}}(0) = 2\pi C\,(1-\tilde{a}_3-\tilde{a}_4) \,,
 \label{f0}
\end{equation}
where the coefficients $\tilde{a}_3$ and $\tilde{a}_4$ are explicitly given by (\ref{tilda_ai}).
\emph{The axis ${\theta=0}$ in the metric (\ref{GP_metr_2}) can thus be made regular} by the unique choice
\begin{align}
 C= C_0 &\equiv\frac{1}{1-\tilde{a}_3-\tilde{a}_4}\,.
 \label{C0}
\end{align}
Notice that for vanishing acceleration this condition is simply ${C_0=1}$ because ${\tilde{a}_3=0=\tilde{a}_4}$ for ${\alpha=0}$, see Section~\ref{gen_no_accel}.

Analogously, we can regularize the \emph{second axis of symmetry} ${\theta=\pi}$. However, there is now a conceptual problem that the metric function $g_{\varphi\varphi}$ (and thus the circumference) does \emph{not} approach zero in the limit ${\theta\to\pi}$ due to the presence of the term ${2\tilde{l}(1-\cos\theta) \to 4\tilde{l}}$. This can be resolved by first applying the transformation of the time coordinate
\begin{equation}
t_{\pi} \equiv t - 4\tilde{l}\,\varphi\,.
 \label{t-tpi}
\end{equation}
The metric (\ref{GP_metr_2}) then becomes
\begin{align}
    \dd s^2=\dfrac{1}{\Omega^2}\Bigg[&-\dfrac{\mathcal{Q}}{\varrho^2}
    \Big(\dd t_{\pi}-\big[a\sin^2\theta
    -2\tilde{l}(1+\cos\theta)\big]\dd\varphi\Big)^2
    +\dfrac{\varrho^2}{\mathcal{Q}}\,\dd \tilde{r}^2 \nonumber\\
    &+\dfrac{\varrho^2}{\tilde{\mathcal{P}}}\,\dd\theta^2+\dfrac{\tilde{\mathcal{P}}}{\varrho^2}\sin^2\theta
    \Big(a\,\dd t_{\pi}-\big[\tilde{r}^{\,2}+(a-\tilde{l}\,)^2\big]\dd\varphi\Big)^2\Bigg],\label{GP_metr_2-regular-pi}
\end{align}
i.e.,
\begin{align}
 g_{\varphi\varphi} &= \frac{1}{\Omega^2\varrho^2}\,
   \Big[\, \tilde{\mathcal{P}}\sin^2\!\theta\, \big[\tilde{r}^{\,2}+(a-\tilde{l}\,)^2\big]^2
   - \mathcal{Q} \big[a\sin^2\theta -2\tilde{l}(1+\cos\theta)\big]^2\,\Big].
    \label{gphiphi-gthetatheta-pi}
\end{align}
For ${\theta \to \pi}$ we thus get ${ g_{\varphi\varphi} \approx \tilde{\mathcal{P}} \big[\tilde{r}^{\,2}+(a-\tilde{l}\,)^2\big]^2\,\dfrac{(\pi-\theta)^2}{\Omega^2\varrho^2}}$.
The radius of a small circle around the axis ${\theta=\pi}$  is ${\int_{\theta}^{\pi}\! \sqrt{g_{\theta\theta}}\,\dd\theta}$, so that the  fraction
\begin{equation}
 f_\pi := \lim_{\theta\to \pi} \frac{\hbox{circumference}}{\hbox{radius}}
 =\lim_{\theta\to\pi} \frac{2\pi C \sqrt{g_{\varphi\varphi}} }{ (\pi-\theta)\,\sqrt{g_{\theta\theta}}}  \,,
 \label{Accel-conpi}
\end{equation}
is
\begin{equation}
 f_\pi =  2\pi C\,\tilde{\mathcal{P}}(\pi) = 2\pi C\,(1+\tilde{a}_3-\tilde{a}_4) \,.
 \label{fpi}
\end{equation}
\emph{The axis ${\theta=\pi}$ in the metric (\ref{GP_metr_2-regular-pi}) is thus regular} for the unique choice
\begin{align}
 C= C_\pi &\equiv \frac{1}{1+\tilde{a}_3-\tilde{a}_4}\,,
 \label{Cpi}
\end{align}
(which for vanishing acceleration $\alpha$ is simply ${C_\pi=1}$).
With such a choice, there is a \emph{deficit angle} $\delta_0$ (conical singularity) along ${\theta=0}$, namely
\begin{eqnarray}
 \delta_0 \equiv 2\pi-f_0 = \frac{4\pi\,\tilde{a}_3}{1+\tilde{a}_3-\tilde{a}_4}\,.
 \label{delta0}
\end{eqnarray}
The tension in the corresponding \emph{cosmic string along ${\theta=0}$ pulls the black hole, causing its uniform acceleration}~$\alpha$.

Complementarily, when the first axis of symmetry ${\theta=0}$ is made regular by the choice (\ref{C0}), there is necessarily an \emph{excess angle} $\delta_\pi$ along the second axis ${\theta=\pi}$, namely
\begin{eqnarray}
 \delta_\pi \equiv 2\pi-f_\pi = -\frac{4\pi\,\tilde{a}_3}{1-\tilde{a}_3-\tilde{a}_4} \,.
 \label{deltapi}
\end{eqnarray}
This represents the \emph{cosmic strut along ${\theta=\pi}$ pushing the black hole}.

Both the axes ${\theta=0}$ and ${\theta=\pi}$ can be \emph{simultaneously regular} if and only if ${\tilde{a}_3=0}$. In the famous C-metric this necessarily requires ${\alpha=0}$. In view of \eqref{tilda_ai}, for our new type of black holes this can be achieved if the physical parameters satisfy the constraint
\begin{equation}
\dfrac{a}{\omega^2}\,(\omega\,a_3+4l\,a_4) = 0 \,.
 \label{bothregularaxes}
\end{equation}

For $l=0,~e=0=g$ it requires $a_3=0$, and using \eqref{prim_param_6-e=0=g,l=0} we obtain the condition $\alpha\, m=0$. This is achieved in the non-acceleration subcases $\alpha=0$, \emph{both for the Kerr black holes}, i.e. \eqref{GP_metric-alpha=0} with $\tilde{l}=0$, but interestingly \emph{also for the Kerr-Bertotti-Robinson black holes} \eqref{P_Kerr-BR}--\eqref{Om2_Kerr-BR} immersed in the external magnetic field \cite{Kerr-BR}.

\subsection{Structure of the electromagnetic field}

To give a more detailed geometric interpretation of the new solution, it is also useful to find the \emph{null eigendirections of the electromagnetic field}, and to clarify their relation to the PNDs of the Weyl tensor. The standard way of finding them is to conduct a null rotation of the PND tetrad \eqref{null_tetr_trans},
\begin{align}\label{null-rotation-fixed-l}
    \mathbf{l}'=\mathbf{l}\,,\qquad
    \mathbf{m}'=\mathbf{m}+K\,\mathbf{l}\,,\qquad
    \mathbf{k}'=\mathbf{k}+K\,\bar{\mathbf{m}}+\bar{K}\,\mathbf{m}+K\bar{K}\,\mathbf{l}\,,
\end{align}
and to find the specific values of $K$ such that the new value of the field component
\begin{align}\label{null-rotation-fixed-l-Phi}
    \Phi_0'=K^2\, \Phi_2+2K\, \Phi_1+\Phi_0
\end{align}
is zero, $\Phi_0'=0$. In general, this equation has 2 complex roots,
\begin{equation}
    K_{\pm}=-\kappa\pm \sqrt{\kappa^2-1}\,,\label{K_pm}
\end{equation}
where the complex parameter $\kappa$ is
\begin{align}
    \kappa :=\Phi_1/\Phi_0\,.
\end{align}
These two values of $K_{\pm}$ identify the two null eigendirections of the electromagnetic field ${\mathbf{k}_{\pm}=\mathbf{k}+K_{\pm}\,\bar{\mathbf{m}}+\bar{K}_{\pm}\,\mathbf{m}+K_{\pm}\bar{K}_{\pm}\,\mathbf{l}}$ of the Maxwell field. In the case $\Phi_0=0$, implying also $\Phi_2=0$, both the Weyl tensor PNDs $\mathbf{k}$ and $\mathbf{l}$ are also the null eigendirections of the electromagnetic field (the fields are aligned as, e.g., for the Kerr-Newman black holes)

It is convenient to introduce alternative quantities describing geometrically how the PNDs of the Weyl tensor and the Faraday null eigendirections differ. Using the orthogonal basis $\mathbf{e}_0=(\mathbf{k}+\mathbf{l})/\sqrt{2}$, $\mathbf{e}_1=(\mathbf{k}-\mathbf{l})/\sqrt{2}$, $\mathbf{e}_3=(\mathbf{m}+\bar{\mathbf{m}})/\sqrt{2}$ and $\mathbf{e}_2=(\mathbf{m}-\bar{\mathbf{m}})/\sqrt{2}\im$, calculated from the null tetrad (\ref{null_tetr_trans}), we express $\mathbf{k}'$ in the form
\begin{equation}
    \mathbf{k}'=\dfrac{1+K\bar{K}}{\sqrt{2}}\,\mathbf{e}_0+\dfrac{1-K\bar{K}}{\sqrt{2}}\,\mathbf{e}_1
        +\im\,\dfrac{\bar{K}-K}{\sqrt{2}}\,\mathbf{e}_2+\dfrac{K+\bar{K}}{\sqrt{2}}\,\mathbf{e}_3\,.
\end{equation}

One obtains a privileged spatial vector $\vec{\mathbf{e}}$ by subtracting the timelike part from $\mathbf{k}'$,
\begin{equation}
    \vec{\mathbf{e}}=\dfrac{1-K\bar{K}}{\sqrt{2}}\,\mathbf{e}_1+\im\,\dfrac{\bar{K}-K}{\sqrt{2}}\,\mathbf{e}_2
        +\dfrac{K+\bar{K}}{\sqrt{2}}\,\mathbf{e}_3\,.
\end{equation}
Its norm is
\begin{equation}
    ||\vec{\mathbf{e}}||=\dfrac{1+K\bar{K}}{\sqrt{2}}\,.
\end{equation}

Now we define a parameter $\delta$ measuring an \emph{angle between $\vec{\mathbf{e}}$ and the privileged spatial vector} $\mathbf{e}_1$, which in our case is the radial vector
\begin{align}
    \mathbf{e}_1=\dfrac{\Omega \sqrt{\mathcal{Q}}}{\varrho}\,\partial_r\,,
\end{align}
and a parameter~$\psi$ measuring an \emph{angle between different polar projections of}  $\Vec{\mathbf{e}}$, namely
\begin{align}
    \cos\delta := \dfrac{\vec{\mathbf{e}}\cdot\mathbf{e}_1}{||\vec{\mathbf{e}}||}=\dfrac{1-K\bar{K}}{1+K\bar{K}}\,,\qquad\qquad
    \psi := \arg K\,.\label{delt_psi_def}
\end{align}

We can prove an important relationship for the two eigendirections roots $K_{\pm}$ given by (\ref{K_pm}), corresponding to the angles $\delta_{\pm}$ and $\psi_{\pm}$, namely that
\begin{align}
    \cos\delta_+=-\cos\delta_-\,,\qquad\qquad\psi_+=2\pi-\psi_-\,.\label{pnds_symm}
\end{align}

Indeed, from \eqref{delt_psi_def} we get
\begin{align}
    \cos\delta_+ +\cos\delta_-=\dfrac{1-K_+\bar{K}_+}{1+K_+\bar{K}_+}+\dfrac{1-K_-\bar{K}_-}{1+K_-\bar{K}_-}
       =2\dfrac{1-K_+\bar{K}_+K_-\bar{K}_-}{(1+K_+\bar{K}_+)(1+K_-\bar{K}_-)}\,.\label{delta+delta-}
\end{align}
Employing specific values of $K_{\pm}$ from \eqref{K_pm}, we infer that
\begin{align}
    K_+K_-\bar{K}_+\bar{K}_-&=\big(\kappa- \sqrt{\kappa^2-1}\big)\big(\kappa+ \sqrt{\kappa^2-1}\big)\big(\bar{\kappa}- \sqrt{\bar{\kappa}^2-1}\big)\big(\bar{\kappa}+ \sqrt{\bar{\kappa}^2-1}\big)\nonumber\\
    &=\Big(\kappa^2- \big(\sqrt{\kappa^2-1}\big)^2\Big)\Big(\bar{\kappa}^2- \big(\sqrt{\bar{\kappa}^2-1}\big)^2\Big)\nonumber\\
    &=1\,,
\end{align}
so that from \eqref{delta+delta-} we indeed obtain $\cos\delta_+ + \cos\delta_-=0$.

This means geometrically that the \textit{two eigendirections of the electromagnetic field are mutually symmetric with respect to a plane spanned by PNDs of the Weyl tensor}. As can be verified, this result is the consequence of the fact that ${\Phi_0=\Phi_2}$, which, in turn, is the consequence of the axial symmetry of the corresponding solution.

\newpage

\begin{figure}[h]
    \centering
\includegraphics[width=1\linewidth]{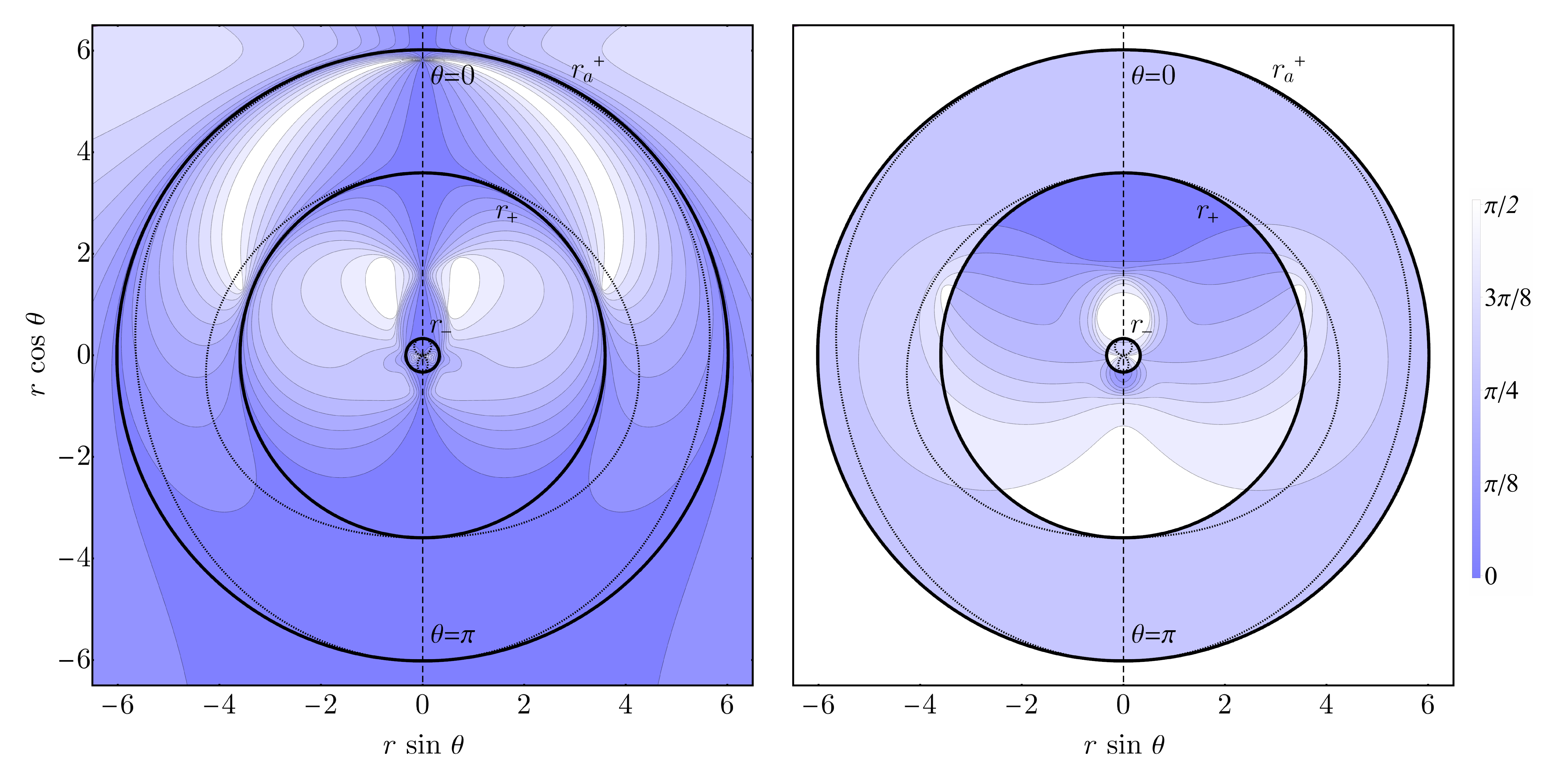}
    \caption{Values of the angles $\delta$ (left panel) and $\psi$ (right panel), defined by (\ref{delt_psi_def}), characterizing how the spacelike part of the eigendirections of the electromagnetic field is rotated relative to the spacelike part of the Weyl tensor PNDs for the \emph{branch~I} of Fig.~\ref{kne_plot}. Black circles represent horizons, dashed curves represent ergoregions. The darkest blue means zero value, so that in such regions with ${\delta=0}$ the fields are aligned.
    The parameters are ${m=2.2,~a=1.1,~l=0.2,~|c|=0.2,~\alpha=0.14,~\omega=1}$.}
    \label{delta_plot_2}
\end{figure}

Let us move to visualization of the electromagnetic field structure, focusing on a solution ${e=0=g}$  (equivalent to ${\beta=\beta_0}$). We can do this simplification because the parameters $e$ and $g$ are related to charges of a black hole itself, while the new feature of this class (namely, the non-alignment of the electromagnetic field) remains even if ${e=0=g}$. For this analysis, we will use the quantities $\delta$ and $\psi$ introduced in (\ref{delt_psi_def}), calculated from \eqref{phi_0-GP}--\eqref{phi_1-GP}. Their dependence on $r$ and $\theta$ is plotted in Fig.~\ref{delta_plot_2} (for the branch I of Fig.~\ref{kne_plot}).

Note that these plots show the angles $\delta$ and $\psi$ for only \textit{one null eigendirection of the electromagnetic  tensor}. Analogous plots for the second eigendirection are not required, as they are symmetric due to Eq.~\eqref{pnds_symm}. As we mentioned in Sec. \ref{sec_em_field}, \textit{on the horizons} and \textit{on the poles} there is $\delta=0$. This means that the corresponding electromagnetic eigendirections \emph{become aligned} with the PNDs of the Weyl tensor there. This was expected because  $\Phi_0$ and $\Phi_2$ given by \eqref{phi_0-GP} are zero where either ${\mathcal{Q}=0}$ or ${\sin\theta=0}$ are zero.

\newpage

\begin{figure}[h]
    \centering
\includegraphics[width=1\linewidth]{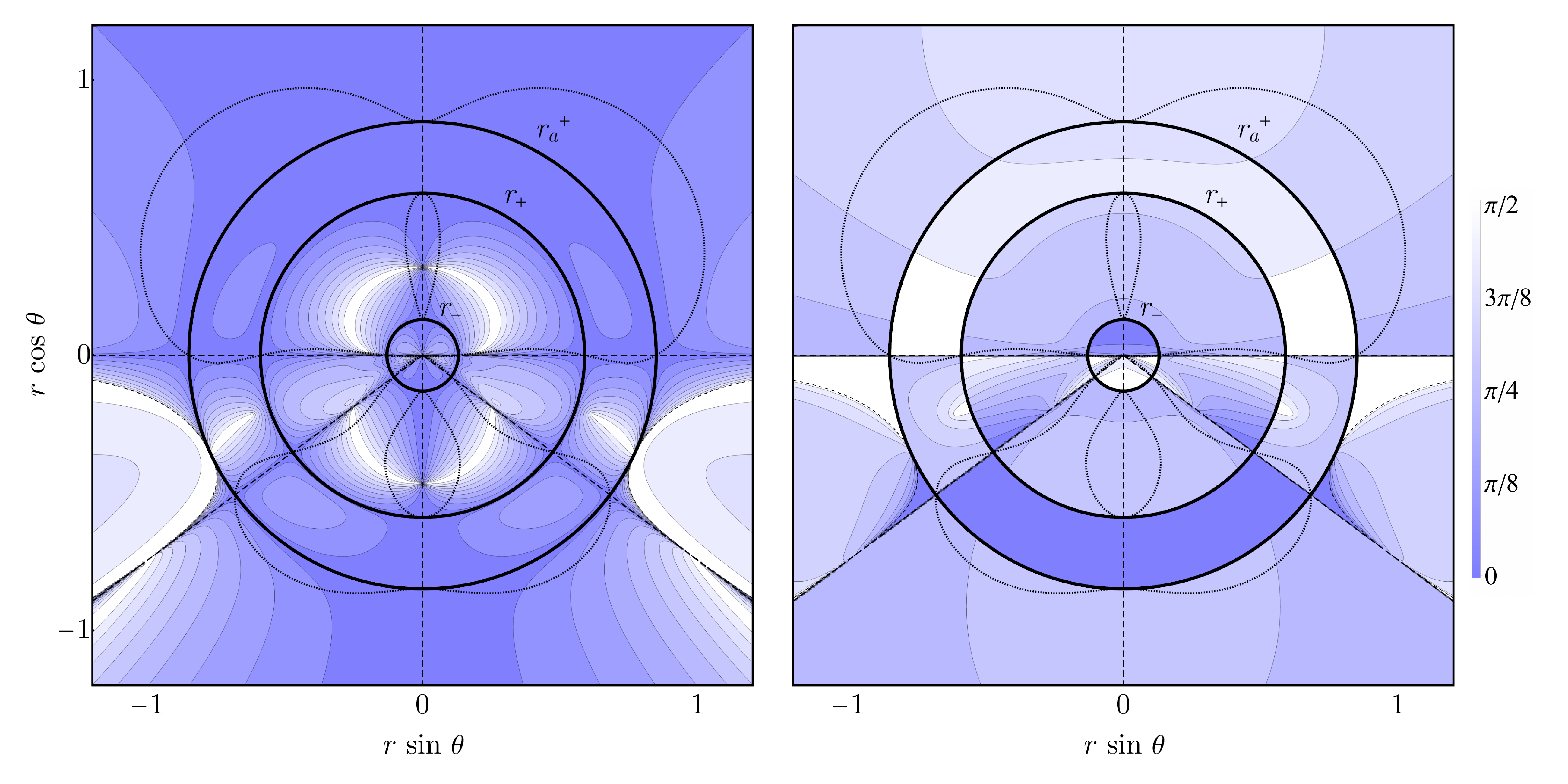}
    \caption{Values of the angles $\delta$ (left panel) and $\psi$ (right panel), defined by (\ref{delt_psi_def}), characterizing how the spacelike part of eigendirections of the electromagnetic field is rotated relative to the spacelike part of the Weyl tensor PNDs for the \emph{branch~III} of Fig.~\ref{kne_plot}. Black circles represent horizons, dashed curves represent ergoregions. The parameters are ${m=2.2,~a=1.1,~l=0.2,~|c|=0.2,~\alpha=0.14,~\omega=1}$.}
    \label{psi_plot_2}
\end{figure}

Such analysis applies to branch~I of Fig.~\ref{kne_plot}. One may also be interested in what happens in other branches, describing new unexpected solutions. For branch III (shown in Fig.~\ref{psi_plot_2}) we observe that it possesses the existence of additional non-trivial zeros of~$\tilde{\mathcal{P}}$ (depicted by thick dashed lines). Such solutions are thus not expected to describe realistic black holes.

\section{Conclusions}

In this work, we constructed a new class of exact twisting solutions to the Einstein-Maxwell equations of algebraic type D with a non-aligned electromagnetic field. In addition to obtaining it by an integration of the field equations, we significantly elaborated on the construction of its several coordinate systems and parameterizations. It turned out that the most useful coordinates for the physical interpretation are the quasipolar ones, in particular (\ref{GP_metr}). The parameters describing this solution are the mass parameter~$m$, the acceleration parameter~$\alpha$, the Kerr parameter~$a$, the NUT parameter~$l$, the ``charge" parameter~$|c|$, and the angles~$\beta$ and~$\gamma$ generating duality rotations between the electric and magnetic parts of the aligned part of the electromagnetic field.

From the analysis of the special cases we have found that the interpretation of some of these parameters is more involved, and admits rich structure of black holes. For example, if $\alpha\rightarrow 0$ while~$|c|$ is kept constant, then one obtains a Kerr-Newman-NUT black hole, and~$c$ represents the value of the charges of a black hole itself. However, if $\alpha\rightarrow 0$ while $\alpha|c|$ is kept constant, one obtains a novel class of black holes immersed in a uniform magnetic field~$B$ (see \cite{Kerr-BR}). In this case, $|c|$ is related to the strengths of the external field. Also, for any nonzero $\alpha$, in the limit $|c|\rightarrow 0$ the electromagnetic field vanishes. This allows us to conclude that in the most general case, $|c|$ is related \textit{solely} to the electromagnetic field, and represents its strength. Also, explicit analysis has shown that the parameter $\gamma$ represents the duality rotation of the whole electromagnetic field, while $\beta$ represents duality rotation between the electric and magnetic charges of a black hole itself.

Even though the parameters $m$, $a$, $l$, and $\alpha$ are related to mass, Kerr and NUT twists, and acceleration, we have to warn that they \textit{are not always equal} to them. A~similar situation exists for the Pleba\'nski-Demia\'nski class, and for various other metric forms, in which the corresponding parameters are different \cite{Ovcharenko2024,Ovcharenko2025}. This is a crucial point, because by setting some parameter to zero (in a given metric form)  does not necessarily erase the corresponding \textit{genuine physical} parameter. A search of various metric forms and reparametrizations is thus still required for this solution. This will allow for a better understanding of the physical properties of the new class.

In addition, we have analyzed various physical properties of the general solution. It turned out that for a given set of $m,~a,~l,~\alpha,~c$ there exist up to 5 different types of black holes with different positions of horizons, and only one of them corresponds to the accelerating Kerr-NUT solutions in the ${c\rightarrow 0}$ limit. This fact may indicate that either several black hole branches exist, or that this is a spurious property of the quasipolar coordinates and specific parametrization we have chosen. This also has to be investigated in future works.

\section*{Supplemental material}

Main expressions and derivations related to this paper are contained in the supplementary Wolfram Mathematica file.

\section*{Acknowledgments}

This work was supported by the Czech Science Foundation Grant No.~GA\v{C}R 23-05914S and by the Charles University Grant No.~GAUK 260325. The authors thank Marcello Ortaggio for valuable discussion concerning the static subcase.

\end{document}